\documentclass[11pt]{article}
\usepackage{rotating} 
\usepackage{adjustbox,lipsum}
\title{Bayesian Inference for Big Spatial Data Using Non-stationary Spectral Simulation}
\author
{Hou-Cheng Yang\footnote{(to whom correspondence should be addressed) Department of Statistics, Florida State University, 117 N. Woodward Ave, Tallahassee, FL 32306, hy15e@my.fsu.edu}\hspace{4pt} and Jonathan R. Bradley\footnote{Department of Statistics, Florida State University, 117 N. Woodward Ave, Tallahassee, FL 32306, bradley@stat.fsu.edu}
	\date{}
}

\usepackage{amsmath,amsthm,bm}
\usepackage{url}
\usepackage{amssymb}
\usepackage{algorithm}
\usepackage{algorithmic}
\usepackage{thmtools}
\usepackage{graphicx,psfrag,epsf}
\usepackage{enumerate}
\usepackage{natbib}
\usepackage[]{hyperref} 
\usepackage{xcolor}

\usepackage{siunitx}
\usepackage{graphics}
\usepackage{epsfig}
\usepackage{booktabs}
\usepackage{multirow}
\usepackage{longtable}
\usepackage{threeparttable}
\begin{document}

\baselineskip24pt

\maketitle

\begin{abstract}
It is increasingly understood that the assumption of stationarity is unrealistic for many spatial processes. In this article, we combine dimension expansion with a spectral method to model big non-stationary spatial fields in a computationally efficient manner. Specifically, we use \cite{mejia1974synthesis}'s spectral simulation approach to simulate a spatial process with a covariogram at locations that have an expanded dimension. We introduce Bayesian hierarchical modelling to dimension expansion, which originally has only been modeled using a method of moments approach. In particular, we simulate from the posterior distribution using a collapsed Gibbs sampler. Our method is both full rank and non-stationary, and can be applied to big spatial data because it does not involve storing and inverting large covariance matrices. Additionally, we have fewer parameters than many other non-stationary spatial models.  We demonstrate the wide applicability of our approach using a simulation study, and an application using ozone data obtained from the National Aeronautics and Space Administration (NASA).

\end{abstract}

\newpage
\section{Introduction}
There is increasing interest in using spatial statistical methods to model environmental processes. This is partially due to the emergence of remote sensing instruments and the popularity of Geographic Information Systems (GIS) software \citep[e.g. see,][for standard references]{stein2006spatial,kalkhan2011spatial}. The main goal of these analyses is to make predictions at observed and unobserved locations and provide uncertainty quantification. Early works make the assumption that the process is weakly stationary \citep[e.g., see][for a review]{cressie1993statistics}; that is, the covariance between the response at two different locations is a function of the spatial lag. However, non-stationary processes are much more common in environmental systems observed over large heterogeneous spatial domains \citep[see][for a disscussion]{bradley2016comparison}. There are many models for non-stationary spatial data, and reduced ranks basis function expansions have become a popular choice \citep{banerjee2008gaussian,cressie2008fixed}. However, there are inferential issues with reduced rank methods in the spatial setting \citep{stein2014limitations}, and consequently, there is renewed interest in proposing computationally efficient full-rank models \citep{nychka2015multiresolution,datta2016hierarchical,katzfuss2017multi,bradleyhierarchical,katzfuss2018vecchia}. Thus, in this article our primary goal is to develop an efficient full rank non-stationary spatial statistical model.

There are numerous methods available to model non-stationary spatial data. For example, process convolution \citep{higdon1998process,paciorek2006spatial,neto2014accounting} convolves a known spatially referenced function with a spatial process typically assumed to be Gaussian. There are several related, but different approaches available. For example, using a finite integral representation of a process convolution results in a basis function expansion \citep[][page 157]{cressie2011statistics}. Several parameterizations of basis function expansions are available, including: fixed rank kriging \citep{cressie2008fixed}, lattice kriging \citep{nychka2015multiresolution}, the predictive process \citep{banerjee2008gaussian}, and a stochastic partial differential equation approach \citep{lindgren2011explicit}, among others.

An alternative to modelling nonstationarity with spatial basis functions is to assume a deformation \citep{sampson1992nonparametric}. Here, Euclidean space is ``deformed," or warped, so that far away locations can be more correlated, and vice versa. The parameter space for this method is considerably smaller than many parameterizations using spatial basis function expansions  \citep[e.g., see][for examples]{cressie2008fixed,kang-cressie-2011}, and is full rank. A similar but different approach to deformation is referred to as ``dimension expansion" \citep{bornn2012modeling}. This method involves extending the dimension of the locations to a higher dimensional space. This methodology is based on the surprising result that every non-stationary covariance function in $\mathbb{R}^{d}$ can be written as a stationary covariogram defined on locations in $\mathbb{R}^{2d}$ \citep{perrin2003nonstationarity}. Recently, \citet{bornn2012modeling} proposed a method of moments approach to analyzing spatio-temporal data using dimension expansion. To our knowledge the dimension expansion approach has not been implemented using a Bayesian framework.

Thus, our first contribution is to introduce dimension expansion to the Bayesian setting to analyze big spatial data. To achieve a computationally efficient approach to dimension expansion in the Bayesian setting we offer three technical results.
In our first technical result, we provide a ``non-stationary version" of Bochner's Theorem \citep{bochner1959lectures}. That is, we show that a non-stationary covariance function can be written as a convolution of the cosine function with a spectral density. The proof of this result simply involves combining \citet{perrin2003nonstationarity}'s dimension expansion result with Bochner's Theorem. This result opens up new opportunities to use spectral methods to model non-stationary spatial process. Other methods exist \citep[e.g. see,][]{priestley1965evolutionary,martin1982time} to model non-stationary data using spectral densities. However, these methods involve difficult to interpret types of ``quasi-stationarity" assumptions \citep[see,][for a discussion]{sayeed1995optimal}, while our approach can be easily interpreted through dimension expansion. \citet{castruccio2017evolutionary} have also proposed an approach that uses evolutionary spectrum and incorporates an axial symmetric structure into their model.


The second technical result developed in this manuscript follows from our non-stationary version of Bochner's Theorem. Specifically, we extend \citet{mejia1974synthesis}'s method for spectral simulation of a stationary spatial processes to non-stationary spatial processes. This makes it straightforward to simulate in the high-dimensional non-stationary setting because spectral simulation does not require the inverse and storage of a high-dimensional covariance matrix (i.e., is matrix free). In practice, Gaussian spatial datasets correspond to a likelihood that is difficult to compute in high dimensions (i.e., when the dimension of the data $n$ is large) because this requires $O(n^{3})$ computation and $O(n^{2})$ dynamic memory. 

Our algorithm is a type of collapsed Gibbs sampler \citep{liu1994collapsed} and it involves two steps. The first step is to augment the likelihood with an $n$-dimensional random vector. Then non-stationary spectral simulation is used within each step of Gibbs sampler to simulate this random vector from it's prior distribution. This strategy is computationally feasible, full-rank, does not require storage of large matrices, and can be implemented on irregularly spaced locations. This last feature is particularly important as spectral methods based on the discrete Fourier transform often require regularly spaced locations \citep{fuentes2002spectral,fuentes2008class}.

The remaining sections of this article are organized as follows. Section 2 introduces our proposed statistical model, and our first two theoretical results. In Section 3, we describe our implementation, which includes inference using a collapsed Gibbs sampler. In Section 4, we present a simulation study, and compare our approach to two different state-of-the art methods in spatial statistics referred to as the Nearest Neighbor Gaussian Process \citep[NNGP;][]{datta2016nearest} and the general Vecchia approximation \citep{katzfuss2017general}. In Section 5, we implement our model using the benchmark ozone dataset analyzed in \citep{cressie2008fixed} and \citep{zhang2018smoothed}. Finally, Section 6 contains a discussion. For ease of exposition all proofs are given in the appendices.
\section{Methodology}
Let $Z(\cdot)$ be a spatial process defined for all $\textbf{s}$$\in$$D$$\subset$$\mathbb{R}^{d}$, where $D$ is the spatial domain of interest in $d$-dimensional Euclidean space, $\mathbb{R}^{d}$. We observe the value of $Z(\cdot)$ at a finite set of locations $\textbf{s}_{1}$, $\ldots$, $\textbf{s}_{n}$$\in$$D$. The data is decomposed additively with
\begin{equation*}
 Z(\textbf{s})=Y(\textbf{s})+\epsilon(\textbf{s}),
\end{equation*}
 where $\textbf{s}$$\in$$D$, $Y(\cdot)$ is the Gaussian process of principal interest, and the Gaussian process $\epsilon(\cdot)$ represents measurement error. The measurement error $\epsilon(\cdot)$ is assumed to be uncorrelated with mean-zero and variance function $V_{\epsilon}(\cdot)=\sigma_{\epsilon}^2\textbf{I}_{n}$.
 
The process $Y(\cdot)$ is further decomposed as 
\begin{equation*}
 Y(\textbf{s})=\textbf{X}^{\prime}(\textbf{s})\pmb{\beta}+\pmb{\nu}(\textbf{s}); \hspace{2pt} \textbf{s}\in D,
\end{equation*}
where $\textbf{X}(s)$ is a known $p$-dimensional vector of covariates, and $\pmb{\beta}\in\mathbb{R}^{p}$ is unknown. For any collection of locations $\textbf{u}_{1}, \ldots, \textbf{u}_{m}$, the random vector $\pmb{\nu}=(\nu(\textbf{u}_{1}), \ldots, \nu(\textbf{u}_{m}))'$ is assumed to have the probability density function (pdf),
\begin{equation}
f(\pmb{\nu}\mid\pmb{\theta})=\int_{\mathbb{R}^{m}}f(\pmb{\nu}\mid\pmb{\theta}, \widetilde{{\pmb{\nu}}},\delta^2)f(\widetilde{\pmb{\nu}}\mid\pmb{\theta})d\widetilde{\pmb{\nu}}, \label{originalcase}
\end{equation}
where $f(\pmb{\nu}\mid\pmb{\theta},\widetilde{\pmb{\nu}},\delta^2)$ is the multivariate normal distribution with mean $\widetilde{\pmb{\nu}}\in\mathbb{R}^{m}$, covariance matrix $\delta^2\textbf{I}_{m}$, and $\textbf{I}_{m}$ is an $m\times m$ identity matrix. The pdf $f(\widetilde{\pmb{\nu}}\mid\pmb{\theta})$ will be specified in Section 2.2, but is approximately normal with mean zero, and covariance matrix $\textit{C}(\pmb{\theta})$, where the $(i,j)$-th element of $\textit{C}(\pmb{\theta})$ is
\begin{equation*}
 C(\textbf{s}_{i},\textbf{s}_{j})=\sigma_{\nu}^{2}\exp\left(-\frac{E(\textbf{s}_i,\textbf{s}_j)}{\phi}\right),
\end{equation*}
 where 
\begin{equation*}
E(\textbf{s}_i,\textbf{s}_j)=\|\bigl(\begin{smallmatrix}
\textbf{s}_{i}\\ \pmb{\psi}^{\prime}(\textbf{s}_{i})\pmb{\eta}
\end{smallmatrix}\bigr)-\bigl(\begin{smallmatrix}
\textbf{s}_{j}\\ \pmb{\psi}^{\prime}(\textbf{s}_{j})\pmb{\eta}
\end{smallmatrix}\bigr)\|,
\end{equation*}
$\pmb{\theta}=(\pmb{\eta}^{\prime},\phi,\sigma_{\nu}^{2})^{\prime}$, and $\|\cdot\|$ is a Euclidean distance. This covariance function uses the aforementioned dimension expansion approach from \cite{bornn2012modeling}. Here, $\pmb{\psi}(\textbf{s}_{i})$ is an $r\times d$ matrix consisting of known basis functions. This use of spatial basis functions is similar to the model in \cite{shand2017modeling}. It will be useful to organize the $n$-dimensional vectors $\textbf{Z}=\{Z(\textbf{s}_{1})\ldots Z(\textbf{s}_{n})\}^{\prime}$, $\textbf{Y}=\{Y(s_{1})\ldots Y(s_{n})\}^{\prime}$ and $\pmb{\epsilon}=\{\epsilon(s_{1})\ldots \epsilon(s_{n})\}^{\prime}$. To model $\textbf{Z}$ set define the prediction locations $\{\textbf{u}_{1}, \ldots, \textbf{u}_{m}\}$ such that the observed locations $\{\textbf{s}_{1}, \ldots, \textbf{s}_{n}\}\subset\{\textbf{u}_{1}, \ldots, \textbf{u}_{m}\}$. Define the corresponding $n\times n$ diagonal matrix $\textbf{V}_{\epsilon}$ $\equiv$ $cov(\pmb{\epsilon})=diag(\text{V}_{\epsilon}(\textbf{s}_{i}): i=1,\ldots,n)$.
\subsection{The Bayesian Hierarchical Model}
In this section, we summarize the statistical model used for inference. The model is organized using the ``data model", ``process model", and ``parameter model" notation used in \cite{cressie2011statistics}, as follows:
\begin{align}
	\nonumber
	&\textbf{Data Model}: \textbf{Z}\mid\pmb{\beta},\pmb{\nu}\sim\text{N}(\textbf{X}\pmb{\beta}+\textbf{O}\pmb{\nu},\sigma_{\epsilon}^2\textbf{I}_{n})\\
	\nonumber
	&\textbf{Process Model 1}: \pmb{\nu}\mid\pmb{\theta},\widetilde{\pmb{\nu}},\delta\sim\text{N}(\widetilde{\pmb{\nu}},\delta^2\textbf{I}_{n})\\
	\nonumber
	&\textbf{Process Model 2}: \widetilde{\pmb{\nu}}\mid\pmb{\theta}\sim f(\widetilde{\pmb{\nu}}\mid\pmb{\theta})\\
	\nonumber
	&\textbf{Parameter Model 1}: \pmb{\beta}\sim\text{N}(\textbf{0},\sigma^{2}_{\beta}\textbf{I}_{p})\\
	\nonumber
	&\textbf{Parameter Model 2}: \pmb{\eta}\sim\text{N}(\textbf{0},\sigma^{2}_{\eta}\textbf{I}_{r})\\
	\nonumber
	&\textbf{Parameter Model 3}: \sigma^{2}_{\nu}\sim \text{IG}(\alpha_{1},\beta_{1})\\
	\nonumber
	&\textbf{Parameter Model 4}: \sigma^{2}_{\beta}\sim\text{IG}(\alpha_{2},\beta_{2})\\
	\nonumber
	&\textbf{Parameter Model 5}: \sigma^{2}_{\eta}\sim\text{IG}(\alpha_{3},\beta_{3})\\
	\nonumber
	&\textbf{Parameter Model 6}: \phi\sim\text{U}(1,\text{U})\\
	\nonumber
	&\textbf{Parameter Model 7}: \delta^{2}\sim\text{IG}(\alpha_{4},\beta_{4}).\\
	&\textbf{Parameter Model 8}: \sigma_{\epsilon}^2\sim\text{IG}(\alpha_{5},\beta_{5}).\label{model}
\end{align}
\noindent
In Equation (\ref{model}), \textbf{O} is an $n\times m$ incidence matrix, $\textbf{0}_{p}$ is a p-dimensional vector of zeros; ``$\text{N}(\pmb{\mu},\pmb{\Sigma})$" is a shorthand for a multivariate normal distribution with mean $\pmb{\mu}$ and positive definite covariance matrix $\pmb{\Sigma}$; ``$\text{IG}(\alpha,\kappa)$" is a shorthand for the inverse gamma distribution with shape $\alpha>0$ and scale $\kappa>0$; and ``$\text{U}(L,U)$" is a shorthand for a uniform distribution with lower bound $L$ and upper bound $U$. All hyperparameters are chosen so that the corresponding prior distribution is ``flat". Example specifications are provided in Section 4 and Section 5. 

Process Model 1, Parameter 1, and Parameter Models 3$\--$5 are fairly standard assumptions for Gaussian data, as they lead to easy to sample full-conditional distributions within a Gibbs sampler \citep{cressie2011statistics}. Parameter model 6 and 7 are used to avoid identifiability issues and leads to a conjugate full-conditional distribution \citep[see][page 124]{banerjee2014hierarchical}. It is common to assume Process Model 2 is $f(\widetilde{\pmb{\nu}}\mid\pmb{\theta})$ is the multivariate normal distribution with mean zero and covariance matrix $\textbf{C}(\pmb{\theta})$ \citep{banerjee2014hierarchical,cressie2011statistics}. However, $f(\widetilde{\pmb{\nu}}\mid\pmb{\theta})$ is only approximately normal with mean-zero and covariance matrix $\textbf{C}(\pmb{\theta})$ (see Section 2.2 for details).
\\
\subsection{Theoretical Considerations}
A non-stationary extension of Bochner's Theorem is stated in Theorem~1.\\

\noindent
\textit{Theorem 1:}\textit{Let $C(\textbf{s}_{i},\textbf{s}_{j})$ be a positive definite function on $D$, which is assumed to be compact and bounded. Then there exists a function $f:D\rightarrow\mathbb{R}^{d}$ and a measure $G(\pmb{\omega})$ such that for any pair of locations $\textbf{s}_{i},\textbf{s}_{j}\in D$,}
\begin{equation}
C(\textbf{s}_{i},\textbf{s}_{j})=\int_{-\infty}^{\infty}\cos[\{\textit{\textbf{f}}(\textbf{s}_{i})-\textit{\textbf{f}}(\textbf{s}_{j})\}^{\prime}\pmb{\omega}_{1}+(\textbf{s}_{i}-\textbf{s}_{j})^{\prime}\pmb{\omega}_{2}]G(d\pmb{\omega}),\label{thm1}
\end{equation}
\textit{where the 2d-dimensional vector} $\pmb{\omega}=(\pmb{\omega}_{1}^{\prime},\pmb{\omega}_{2}^{\prime})^{\prime}$.\\

\noindent
\textit{Proof}: See Appendix A.\\

\noindent
The proof of Theorem 1 involves a simple combination of the result in \citet{perrin2003nonstationarity} and Bochner's Theorem. We use Theorem 1 to define a nonstationary covariance function $C(\cdot,\cdot)$. That is, in our model we choose a specific form for $G(d\omega)$ and $f(\cdot)$, and we use Equation \eqref{thm1} to define our nonstationary covariance function.

Additionally, in practice we assume $\textbf{\textit{f}}(\cdot)=\pmb{\psi}^{\prime}(\textbf{s})\pmb{\eta}$, which is similar to the strategy used in \citet{bornn2012modeling} and \citet{shand2017modeling}. This leads naturally to questions on how to specify spatial basis functions. In general, we use radial basis functions with equally spaced knot locations as suggested in \citet{nychka} and \citet{cressie2008fixed}. One might also consider the use of information criteria to adaptively select knot locations \citep{bradley2011,TzengHuang}. 

There are several things we can learn from Theorem 1. First, every non-stationary covariance function can be written as a convolution in 2\textit{d}-dimensional space according to \eqref{thm1}. Second, $\{\textit{\textbf{f}}(\textbf{s}_{i})-\textit{\textbf{f}}(\textbf{s}_{j})\}^{\prime}\pmb{\omega}_{1}$ is a deformation, which shows an explicit connection between dimension expansion and deformation. Furthermore, this deformation induces non-stationarity, since $\{\textit{\textbf{f}}(\textbf{s}_{i})-\textit{\textbf{f}}(\textbf{s}_{j})\}^{\prime}\pmb{\omega}_{1}=0$ leads to the classical version of Bochner's Theorem. Third, if we assume a specific form of $G(d\pmb{\omega})$, we can use Equation (\ref{thm1}) to approximate the covariance function. For example, when $C(\cdot,\cdot)$ is the exponential covariance function (as is the case in \eqref{model}), then $G(d\pmb{\omega})$ has a corresponding Cauchy density \citep{steinML}. We provide a discussion and consider several other covariograms. Our empirical results suggest that the results appear robust to the choice of covariogram. Denote the density corresponding to $G(d\pmb{\omega})$ with $\frac{g(\omega)}{C(0,0)}$. Moreover, the ability to simulate from the spectral density without mathematical operations of covariance matries, allows us to completely circumvent computing and storing a covariance matrix \citep{mejia1974synthesis}.
\\

\noindent
\textit{Theorem 2:}
\textit{Let $\pmb{\omega}_{i} = (\pmb{\omega}_{1,i}^{\prime}, \pmb{\omega}_{2,i}^{\prime})^{\prime}$, $\pmb{\omega}_{i}\overset{ind}{\sim} G(d\pmb{\omega})$, and $\kappa_i\overset{ind}{\sim}U(-\pi,\pi)$. Then for a given $f:\mathbb{R}^{d}\to\mathbb{R}^{d}$ the random process,}
\begin{equation}
\widetilde{\nu}(s)\equiv\sigma_{\nu}\left(\frac{2}{K}\right)^{\frac{1}{2}}\sum_{i=1}^{\text{K}}\cos(\textbf{\textit{f}}(\textbf{s})^{\prime}\pmb{\omega}_{1,i}+\textbf{s}^{\prime}\pmb{\omega}_{2,i}+\kappa_i),\label{3}
\end{equation}
\textit{$E\{\widetilde{\nu}(s)\}=0$, and $E\{\widetilde{\nu}(\textbf{s}_{i})\widetilde{\nu}(\textbf{s}_{j})\}=C(\textbf{s}_{i},\textbf{s}_{j})$, and converges in distribution (as $K\to\infty$) to a mean-zero Gaussian process with covariance $C(\textbf{s}_{i},\textbf{s}_{j})$ in Equation \eqref{thm1} with spectral density  $\prod \frac{g(\omega_{jk})}{\textit{C}(0,0)}$.} 
\\

\noindent
\textit{Proof}: See Appendix A.\\

\noindent
The proof of Theorem 2 involves a simple combination of the result in \citet{perrin2003nonstationarity} and \citet[][pg. 204]{cressie1993statistics}. In practice, to use Theorem 2, we need to specify the spectral density. In our implementation, we assume that $\omega_{j,i} \overset{ind}{\sim} \frac{g(\omega)}{C(0,0)}$, where for each $i$, $\pmb{\omega}_{i} = (\omega_{1,i},\ldots, \omega_{2d,i})^{\prime}$ and $g(\cdot)$ is the Cauchy density. This choice of the Cauchy density leads to the exponential covariogram \citep{cressie1993statistics}.

It is arguably more common to simulate $\pmb{\nu}$ using a Cholesky decomposition. However, this requires order $n^{3}$ computation and order $n^{2}$ memory. Theorem 2 allows us to simulate $\pmb{\nu}$ without these memory and computational problems. It follows from the transformation theorem \citep{resnick2013probability} that the pdf of $\widetilde{\pmb{\nu}}$ is given, under our specification, by
\begin{align}
\nonumber
&f(\widetilde{\pmb{\nu}}\mid\pmb{\theta})\\
&=\int_{\widetilde{\pmb{\nu}}:\widetilde{\nu}(s_i)=\sigma_{\nu}\left(\frac{2}{K}\right)^{\frac{1}{2}}\sum_{i=1}^{\text{K}}\cos(\textbf{\textit{f}}(s_i)^{\prime}\pmb{\omega}_{1,i}+s_i^{\prime}\pmb{\omega}_{2,i}+\kappa_i)\}}\prod_{jk} \frac{g(\omega_{j,k})}{\textit{C}(0,0)}\prod_{i=1}^{n}\frac{1}{2\pi}I(-\pi<\kappa<\pi)d\pmb{\omega}_{1,i}d\pmb{\omega}_{2,i}d\kappa_{1}\ldots d\kappa_{n},
\label{thm333}
\end{align}
where $I(\cdot)$ is the indicator function. Again, from \citet[][pg. 204]{cressie1993statistics} and Theorem 2 the pdf in \eqref{thm333} is roughly Gaussian with mean zero and covariance $C(\pmb{\theta})$. 
\section{Collapsed Gibbs Sampling}
In this section, we outline the steps needed for collapsed Gibbs sampling. Gibbs sampling requires simulating from full-conditional distributions \citep{gelfand1990sampling}. In a collapsed Gibbs sampler, some of the events conditioned on in the full-conditional distribution are integrated out \citep{liu1994collapsed}. For simplicity, we use the bracket notation, where [$X\mid Y$] represents the conditional distribution of a \textit{X} given \textit{Y} for generic random variables \textit{X} and \textit{Y}. In Algorithm 1, we present the steps needed for our proposed collapsed Gibbs sampler.
\begin{algorithm}[H]
	\caption{Implementation: Collapsed Gibbs sampler}
	\begin{algorithmic}[1]
		\STATE Initialize ${\pmb{\beta}}^{[1]}$, ${\pmb{\nu}}^{[1]}$,
		$\pmb{\eta}^{[1]}$,$\sigma_{\nu}^{2[1]}$,$\sigma_{\beta}^{2[1]}$,
		$\sigma_{\eta}^{2[1]}$, $\phi^{[1]}$ and $\delta^{2[1]}$
		\STATE Set \textit{b} = 2.
		\STATE Simulate ${\pmb{\beta}}^{[b]}$ from $[{\pmb{\beta}}\vert {\pmb{\nu}}^{[b-1]}, \pmb{\eta}^{[b-1]}, \sigma_{\nu}^{2[b-1]}, \sigma_{\beta}^{2[b-1]}, \sigma_{\eta}^{2[b-1]},\phi^{[b-1]},\delta^{2[b-1]},\textbf{Z}]$. 
		\STATE Simulate $\widetilde{\pmb{\nu}}$ from $f(\widetilde{\pmb{\nu}}\mid\pmb{\theta})$ using Theorem 2 with $K$ ``large".
		\STATE Simulate ${\pmb{\nu}}^{[b]}$ from $[{\pmb{\nu}}\vert {\pmb{\beta}}^{[b]}, \pmb{\eta}^{[b-1]}, \sigma_{\nu}^{2[b-1]}, \sigma_{\beta}^{2[b-1]}, \sigma_{\eta}^{2[b-1]},\phi^{[b-1]},\delta^{2[b-1]},\{\widetilde{\pmb{\nu}}\},\textbf{Z}]$.	    
		\STATE Simulate ${\pmb{\eta}}^{[b]}$ from $[{\pmb{\eta}}\vert {\pmb{\beta}}^{[b]}, {\pmb{\nu}}_{m}^{[b]}, \sigma_{\nu}^{2[b-1]}, \sigma_{\beta}^{2[b-1]}, \sigma_{\eta}^{2[b-1]},\phi^{[b-1]},\delta^{2[b-1]}, \textbf{Z}]$, ${\pmb{\nu}}_{m}^{[b]}$ is $m$-dimensional ($m<n$) consisting of $m$ distinct elements of ${\pmb{\nu}}^{[b]}$.
		\STATE Simulate ${\sigma}_{\nu}^{2[b]}$ from $[{\sigma}_{\nu}^2\vert {\pmb{\beta}}^{[b]}, {\pmb{\nu}}^{[b]}, \pmb{\eta}^{[b]}, \sigma_{\beta}^{2[b-1]}, \sigma_{\eta}^{2[b-1]},\phi^{[b-1]},\delta^{2[b-1]},\{\widetilde{\pmb{\nu}}\},\textbf{Z}]$.
		\STATE Simulate ${\sigma}_{\beta}^{2[b]}$ from $[{\sigma}_{\beta}^2\vert {\pmb{\beta}}^{[b]}, {\pmb{\nu}}^{[b]}, \pmb{\eta}^{[b]}, \sigma_{\nu}^{2[b]}, \sigma_{\eta}^{2[b-1]},\phi^{[b-1]},\delta^{2[b-1]},\textbf{Z}]$.   
		\STATE Simulate ${\sigma}_{\eta}^{2[b]}$ from $[{\sigma}_{\eta}^2\vert {\pmb{\beta}}^{[b]}, {\pmb{\nu}}^{[b]}, \pmb{\eta}^{[b]}, \sigma_{\nu}^{2[b]}, \sigma_{\beta}^{2[b]},\phi^{[b-1]},\delta^{2[b-1]},\textbf{Z}]$.
		\STATE Simulate ${\phi}^{[b]}$ from $[\phi\vert {\pmb{\beta}}^{[b]}, {\pmb{\nu}}_{m}^{[b]}, \pmb{\eta}^{[b]}, \sigma_{\nu}^{2[b]}, \sigma_{\beta}^{2[b]},\sigma_{\eta}^{2[b]},\delta^{2[b-1]}, \textbf{Z}]$.
		\STATE Simulate ${\delta}^{2[b]}$ from $[\delta^2\vert {\pmb{\beta}}^{[b]}, {\pmb{\nu}}^{[b]}, \pmb{\eta}^{[b]}, \sigma_{\nu}^{2[b]}, \sigma_{\beta}^{2[b]},\sigma_{\eta}^{2[b]},\phi^{[b]},\{\widetilde{\pmb{\nu}}\},\textbf{Z}]$.
		\STATE Let $b = b+1$.
		\STATE If $b < B$ (a prespecified value) repeat Steps 3 $\--$ 12, otherwise stop.
	\end{algorithmic}
\end{algorithm}
\noindent The expressions for the full-conditional distributions listed in Algorithm 1 are derived in Appendix B. This collapsed Gibbs sampler can easily be modified to allow for heterogeneous variances, and allow for other choices for prior distributions.

The main motivation for collapsed Gibbs sampling is that Step 4 of Algorithm 1 is computationally straightforward. Additionally, in Step 5, the full-conditional distribution has a known, and easy to sample from expression. This is significant, as this full-conditional distribution traditionally involves inverses and determinants of high-dimensional matrices. Specifically, the following relationship holds,
\begin{equation*}
f(\pmb{\nu}\vert\cdot)\propto\exp\left\lbrace-\frac{(\textbf{Z}-\textbf{X}\pmb{\beta}-\textbf{O}\pmb{\nu})^{\prime}\textbf{V}_{\epsilon}^{-1}(\textbf{Z}-\textbf{X}\pmb{\beta}-\textbf{O}\pmb{\nu})}{2}\right\rbrace f(\pmb{\nu}\vert\widetilde{\pmb{\nu}},\delta,\pmb{\theta}),
\end{equation*}
where $ f(\pmb{\nu}\vert\hat{\pmb{\nu}}, \delta,\pmb{\theta})=\exp\left\lbrace-\frac{(\pmb{\nu}-\widetilde{\pmb{\nu}})^{\prime}(\pmb{\nu}-\widetilde{\pmb{\nu}})}{2\delta^{2}}\right\rbrace$. Then,
\begin{equation*}
f(\pmb{\nu}\vert\cdot)\propto\exp\left\lbrace-\frac{\pmb{\nu}^{\prime}(\delta^{-2}\textbf{I}+\textbf{O}'\textbf{V}_{\epsilon}^{-1}\textbf{O})\pmb{\nu}}{2}+\nu^{\prime}\left(\textbf{O}'\textbf{V}_{\epsilon}^{-1}(\textbf{Z}-\textbf{X}\pmb{\beta})+\frac{1}{\delta^{2}}\widetilde{\pmb{\nu}}\right)\right\rbrace.
\end{equation*}
This gives
\begin{equation*}
f(\pmb{\nu}\vert\cdot)=\text{N}(\pmb{\mu}^{*},\pmb{\Sigma}^{*}),
\end{equation*}
where 	$\pmb{\mu}^{*}=\pmb{\Sigma}^{*}\{\textbf{O}'\textbf{V}_{\epsilon}^{-1}(\textbf{Z}-\textbf{X}\pmb{\beta})+\frac{1}{\delta^{2}}\widetilde{\pmb{\nu}}\}$, and $(\pmb{\Sigma}^{*})^{-1}=\delta^{-2}\textbf{I}+\textbf{O}'\textbf{V}_{\epsilon}^{-1}\textbf{O}$, where we emphasis that $\pmb{\Sigma}^{*}$ is a computationally advantageous diagonal matrix.

\section{Simulation Studies}
We simulate data in a variety of settings, and compare to the current state-of-the-art in spatial statistics, the nearest neighbor Gaussian process (NNGP) model \citep{datta2016nearest} and the general Vecchia approximation \citep{katzfuss2017general,katzfuss2018vecchia}. The data are generated in several different ways, all of which differs from the model we fit. That is, we assume $\textbf{Z}=\textbf{Y}+\pmb{\epsilon}$, where $\textbf{Y}$ is a fixed and known $n$-dimensional vector and $\pmb{\epsilon}\sim N(0,\sigma_{\epsilon}^2\textbf{I}_n)$. We choose $\sigma_{\epsilon}^2$ based on the signal-noise-ratios (SNR equal to 2, 3, 5, and 10). For each SNR we allow for three different missing data assumptions. Specifically, 5\% missing at random, 10\% missing at random, and 20\% missing at random. In total, this produces 60 settings. 

\begin{figure}[H]
	\centering
	\includegraphics[width=12cm]{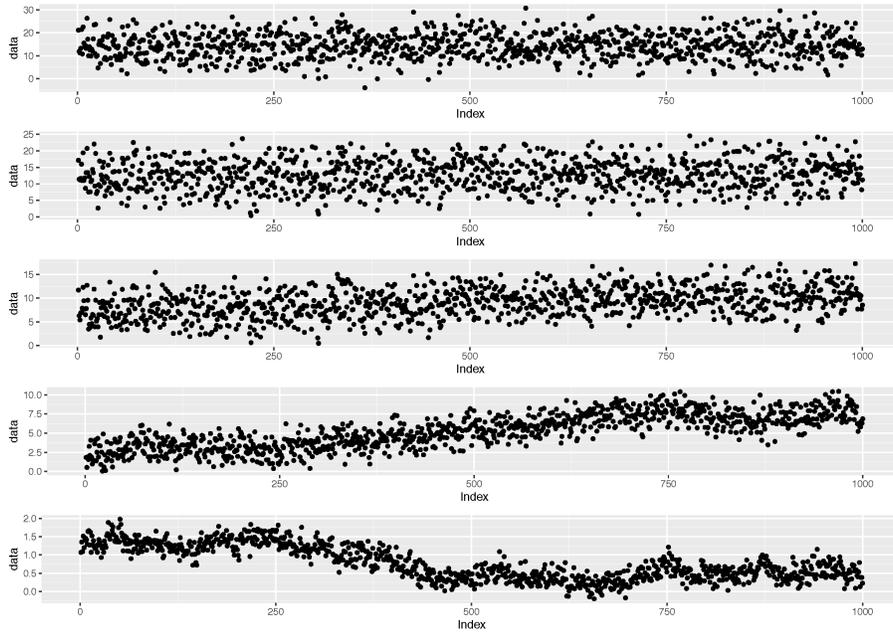}
	\caption{Simulation data with SNR=5; First row to last row are examples of Case 1 to Case 5.}\label{DataSNR5}
\end{figure}

To define $f(\cdot)$ we use a nonlinear function proposed by \citet{friedman1991multivariate}, where for $\bf{x}=(x_1,\ldots, x_5)'$:
\begin{align*}
&f_0(\textbf{x})=10\text{sin}(\pi x_1 x_2)+20(x_3-0.5)^2+10x_4+5x_5,
\end{align*}
 which implies that far away observations may be less similar, suggesting nonstationarity. Then we propose five simulation cases,
\begin{align*}
&\text{Case 1:}\quad Y(i)= f_0(\textbf{x}_i)\\
&\text{Case 2:}\quad Y(i)= 0.85f_0(\textbf{x}_i)+0.15\pmb{\zeta}(i)\\
&\text{Case 3:}\quad Y(i)= 0.5f_0(\textbf{x}_i)+0.5\pmb{\zeta}(i)\\
&\text{Case 4:}\quad Y(i)= 0.15f_0(\textbf{x}_i)+0.85\pmb{\zeta}(i)\\
&\text{Case 5:}\quad Y(i)= \pmb{\zeta}(i)
\end{align*}
and $\pmb{\zeta}=(\pmb{\zeta}(1),\ldots\pmb{\zeta}(n))'\sim\text{N}(0,R(\theta))$, where the $n\times n$ matrix  $R(\theta)=\{\sigma_{\zeta}^{2}\exp\left(-\phi_{\zeta}\|i-j\|\right)\}$. So, the $\pmb{\zeta}$ is a stationary term. In Case 1, the data is generated from a highly nonstationary process and $\bf{x}_i$ is a five dimensional vector consisting of independent draws from a uniform distribution. In Case 3, we weight the process half with a nonlinear term and half with a stationary term. In Case 5, we only have a stationary term. So the data is rough in Case 1 and gradually becomes smoother as we consider other cases. We show examples of the data in Figure \eqref{DataSNR5}.

We generate 1,000 observations over this one-dimensional domain $\left[0,1\right]$ for each case. For Case 2 to Case 4, $\sigma^2_{\zeta}$ is set to be equal to the sample variance of the elements in the vector $(f(\textbf{x}_1),\ldots, f(\textbf{x}_n))'$. We fixed $\phi_{\zeta}=0.3$. SNR is defined to be
\begin{equation*}
	\text{SNR}=\frac{\sum_{i=1}^{n}(Y(i)-\frac{1}{n}\sum_{j=1}^{n}Y(j))^2}{(n-1)\sigma_{\epsilon}^2}
\end{equation*}
so that
\begin{equation*}
	\sigma_{\epsilon}^2=\frac{\sum_{i=1}^{n}(Y(i)-\frac{1}{n}\sum_{j=1}^{n}Y(j))^2}{(n-1)\text{SNR}}.
\end{equation*}
 Each SNR has five different cases as we mention above. In practice, we often do not observe all the useful covariates. Thus, when implementing methods for spatial prediction, we only use $X=(x_1,x_3,x_4,x_5)$, which removes $x_2$. 

In our model, the 20-dimensional vector $\pmb{\psi}(i)$ was chosen to consist of Gaussian radial basis functions over equally spaced knots. That is, $\pmb{\psi}(i)=(\psi_{1}(i)\ldots \psi_{20}(i))^{\prime}$, where 
\begin{equation*}
\psi_{k}(i)=\exp(-\tau\|i-{c}_{k}\|);\hspace{2pt}i = 1,\ldots, 1,000, \hspace{2pt} k=1,\ldots,20,
\end{equation*}
and $\{{c}_{1},\ldots,{c}_{20}\}$ are the equally-spaced knots points over $\{1,\ldots,1,000\}$, and $\tau$ is equal to 1.5 times the median of non-zero distances between the points in $\{1,\ldots,1,000\}$. We set the spectral density equal to
\begin{equation*}
g(\omega)=\frac{1}{\pi(1+\omega^2)}.
\end{equation*}
In Table \eqref{Table:Estimation}, we provide the root mean squared prediction error (RMSPE) of each model for the data in the first row in Figure \eqref{DataSNR5}. The RMSPE is defined as
\begin{equation}
\sqrt{\frac{1}{1,000}\left\lbrace\sum_{i=1}^{1,000}\{Y(i)-\hat{Y}(i)\}^{2}\right\rbrace},\label{mspe}
\end{equation}
where $\hat{Y}(i)$ is the posterior mean from fitting the each model. Here we see that our method (referred to as the Expanded Spectral Density (ESD) method) clearly outperform NNGP and the Vecchia approximation. Our model performs well in terms of estimation as well. Using the data in the first row of Figure \eqref{DataSNR5}, and a half-t prior for $\sigma_{\epsilon}$, we have a posterior mean of $\sigma_{\epsilon}^2$ equal to 5.79 and highest posterior density interval, (2.68, 9.96). The true value is 4.813, which is contained in the interval.

\begin{table}[H]
	\begin{center}
		\caption{Estimation of $\sigma_{\epsilon}^2$ for the data in the first row of Figure \eqref{DataSNR5}. We call our method the Expanded Spectral Density (ESD) method.}
		\label{Table:Estimation}
		\begin{tabular}{c|c} 
			\hline
			\textbf{Method}  &\textbf{RMSPE}  \\
			\hline
			\textbf{ESD} &1.94  \\
			\textbf{NNGP} &3.24  \\
			\textbf{Vecchia Approximation} &3.05  \\
			\hline
		\end{tabular}
	\end{center}
\end{table}

\subsection{Comparisons Over Multiple Replicates}
\begin{figure}[H]
	\centering
	\includegraphics[width=12cm]{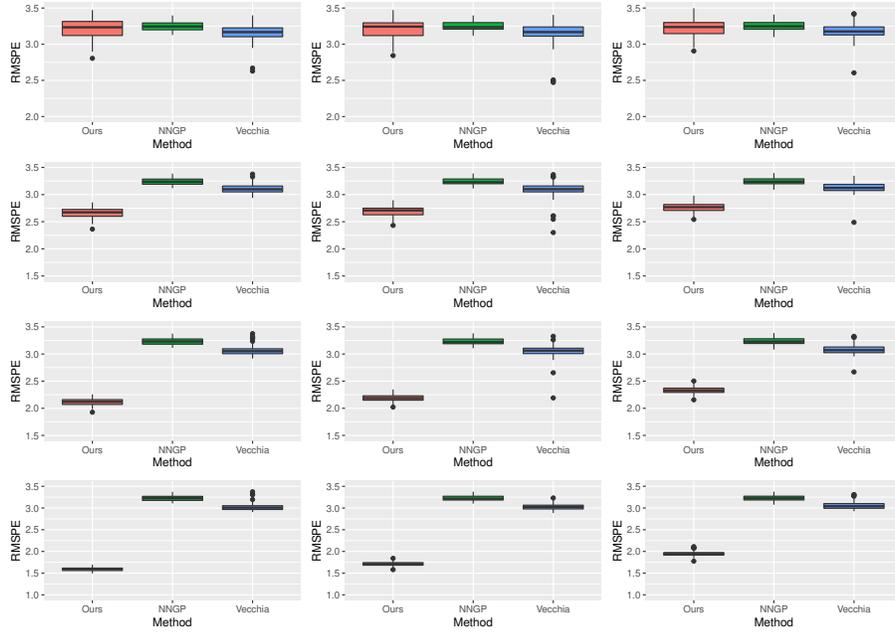}
	\caption{RMSPE for Case 1. First column to last column are results over 5\%, 10\% and 20\% missing at random. First row to last row are example of SNR equal to 2,3,5 and 10.}\label{mspeplotS1}
\end{figure}

\begin{figure}[H]
	\centering
	\includegraphics[width=12cm]{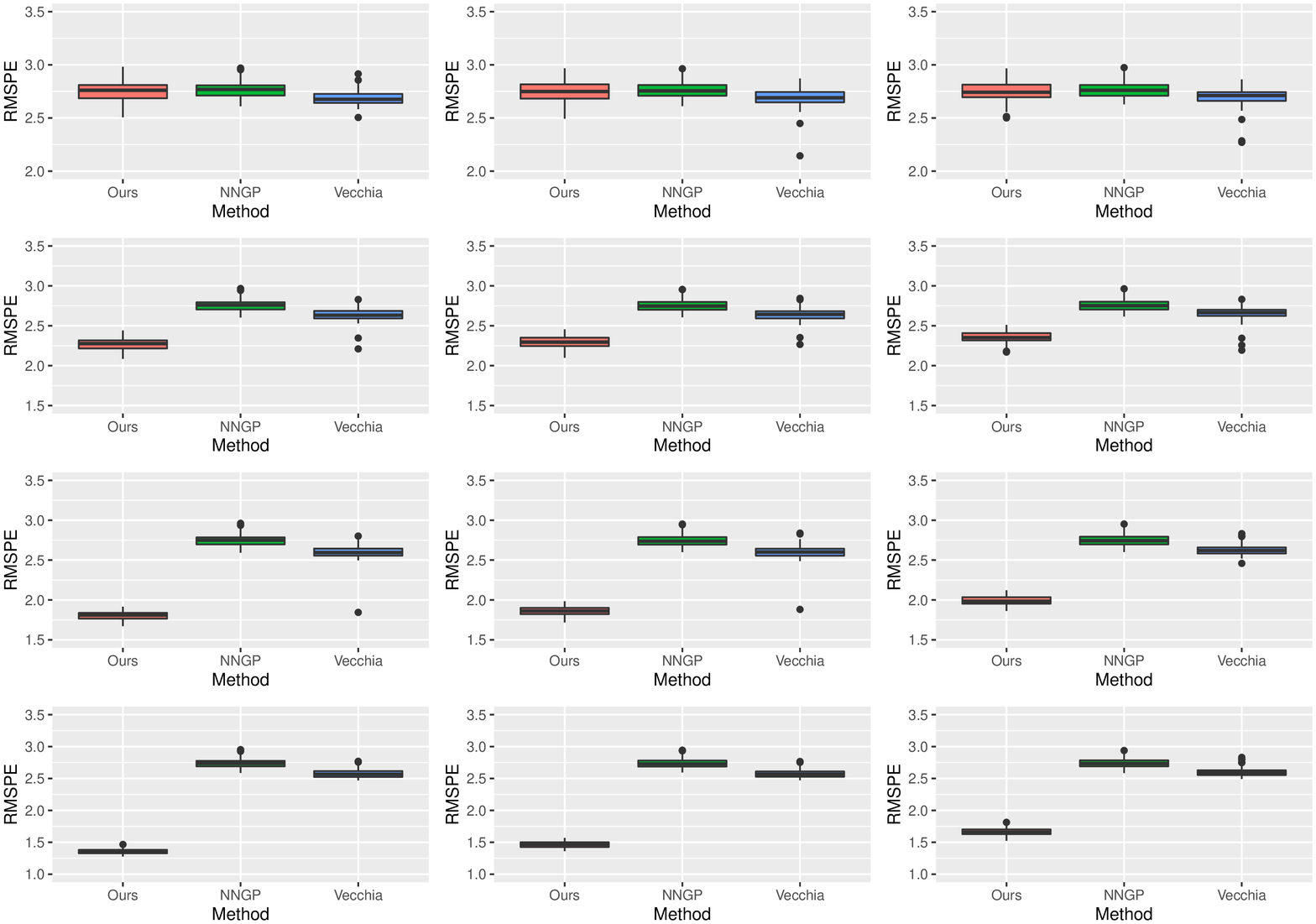}
	\caption{RMSPE for Case 2. First column to last column are results over 5\%, 10\% and 20\% missing at random. First row to last row are example of SNR equal to 2,3,5 and 10.}\label{mspeplotS2}
\end{figure}

\begin{figure}[H]
	\centering
	\includegraphics[width=12cm]{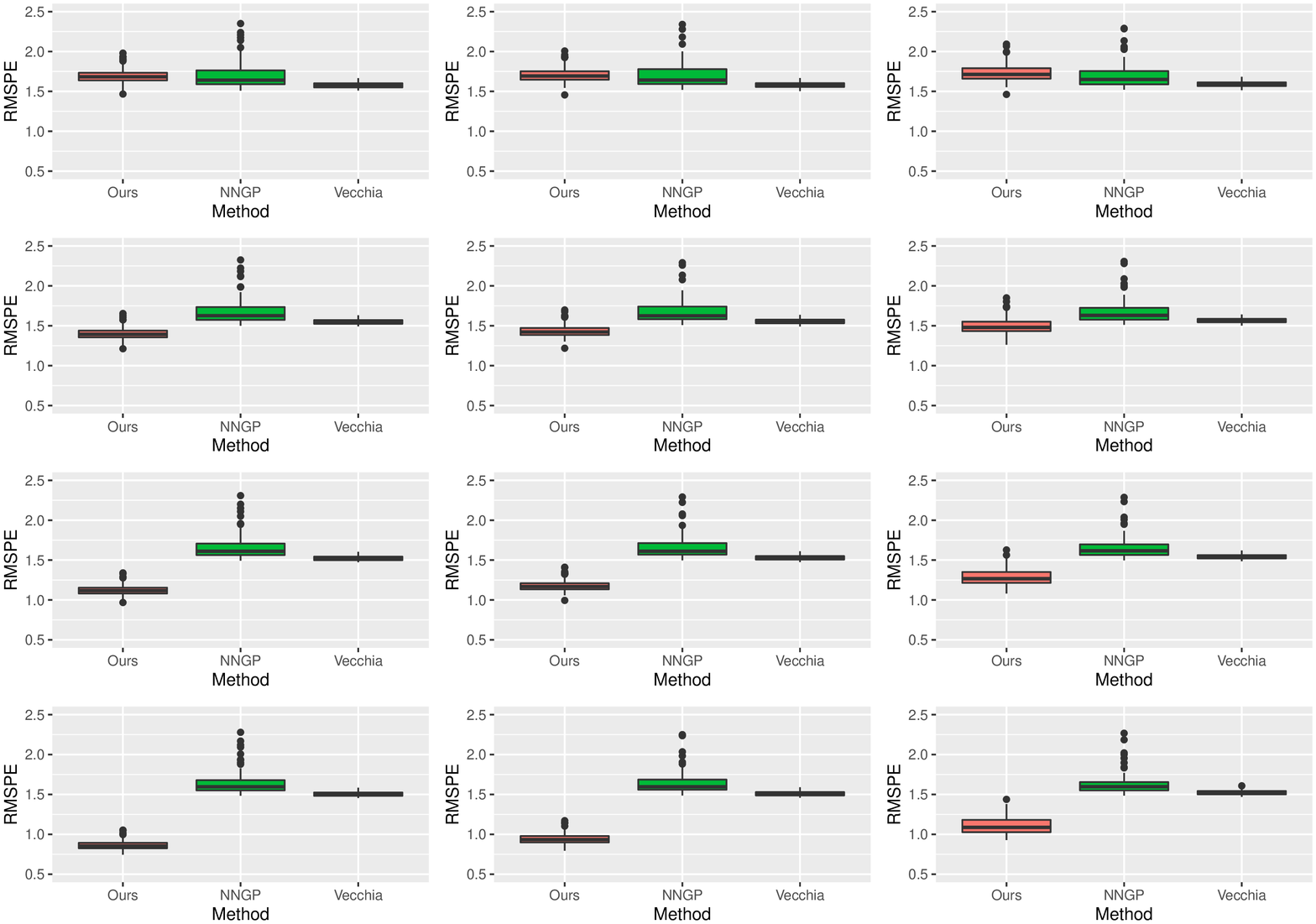}
	\caption{RMSPE for Case 3. First column to last column are results over 5\%, 10\% and 20\% missing at random. First row to last row are example of SNR equal to 2,3,5 and 10.}\label{mspeplotS3}
\end{figure}

\begin{figure}[H]
	\centering
	\includegraphics[width=12cm]{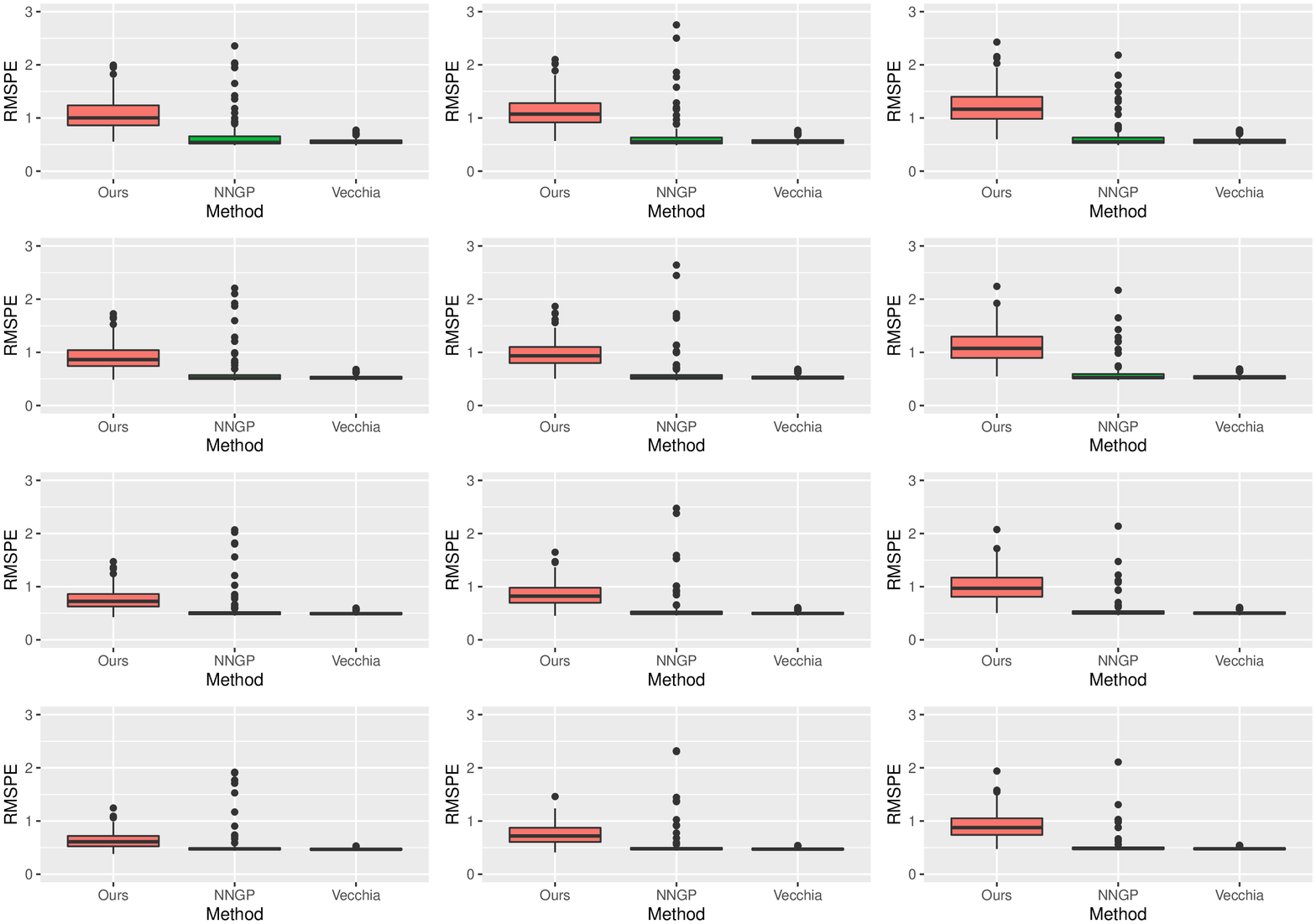}
	\caption{RMSPE for Case 4. First column to last column are results over 5\%, 10\% and 20\% missing at random. First row to last row are example of SNR equal to 2,3,5 and 10.}\label{mspeplotS4}
\end{figure}

\begin{figure}[H]
	\centering
	\includegraphics[width=12cm]{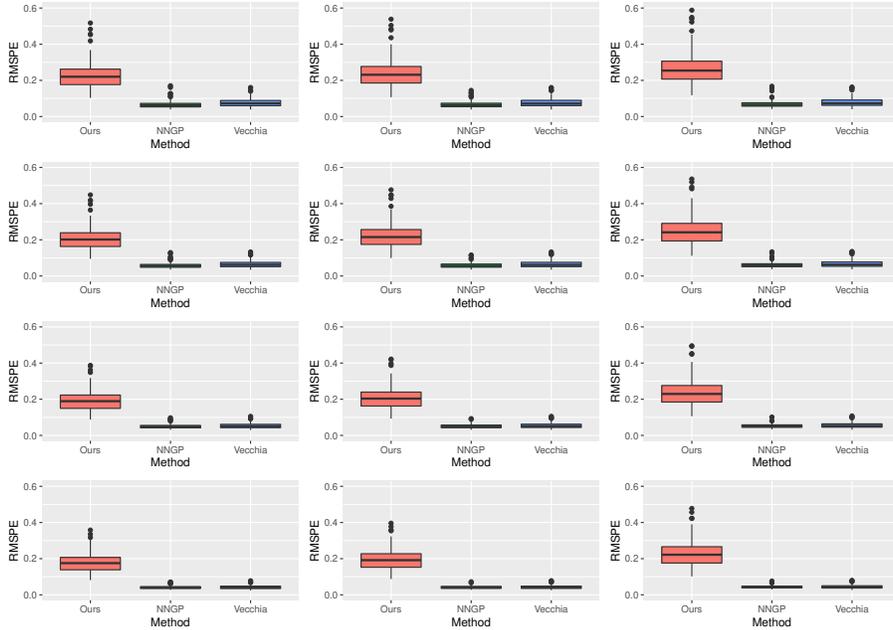}
	\caption{RMSPE for Case 5. First column to last column are results over 5\%, 10\% and 20\% missing at random. First row to last row are example of SNR equal to 2,3,5 and 10.}\label{mspeplotS5}
\end{figure}

To implement NNGP, we use 15 nearest neighbors which is consistent with what is suggested in \citet{datta2016hierarchical}. For the general Vecchia approximation, we use 15 nearest neighbors which is the same as NNGP. We also use the R-package \texttt{spNNGP} \citep{spNNGP} and \texttt{GpGp} \citep{guinness2018gpgp}.  We record the performance (in terms of RMSPE) over the signal noise ratio (SNR) and the proportion of missing in the datasets for each case. The results are show in Figure \eqref{mspeplotS1} to Figure \eqref{mspeplotS5} for Case 1 to Case 5, respectively. For each figure, the first row is for SNR equal 2, the second row if for SNR equal 3, the third row is for SNR equal 5 and the fourth row is for SNR equal 10. For each row, the left plot has 5\% of the data missing, the middle plot has 10\% of the data missing and the right plot has 20\% of the data missing. The boxplot is computed over 100 independent replicates of the one thousand dimensional dataset. We compare each simulation with NNGP and the general Vecchia approximation. 

For both Case 1 and Case 2, we find that our method outperforms the NNGP and the general Vecchia approximation when SNR equals 3, 5 and 10. However, we have a similar result with NNGP and perform slight worse than general Vecchia approximation when SNR=2. In Case 3, we find that our method outperforms the NNGP and general Vecchia approximation when SNR equals 5 and 10, and only slightly outperforms NNGP and general Vecchia approximation when SNR=3. Additionally, ESD performs slightly worse than general Vecchia approximation when SNR=2 and we are in Case 3. In Case 4, we find that general Vecchia approximation outperforms our method, and our method outperforms than NNGP in a few replicates. In Case 5, the stationary case, general Vecchia approximation and NNGP outperform our method, which is not surprising considering that our method is derived for nonstationary processes. Based on the results, we believe our model performs well in highly nonlinear setting with moderate to high signal-to-noise ratios. However, we find the RMSPE is worse than NNGP and general Vecchia approximation when as the process becomes smoother.

%

\section{Real Data Application}
\subsection{Ozone Data Application: Data Description}
\begin{figure}[H] 
	\begin{center}
		\begin{tabular}{c}
			\includegraphics[width=.45\textwidth]{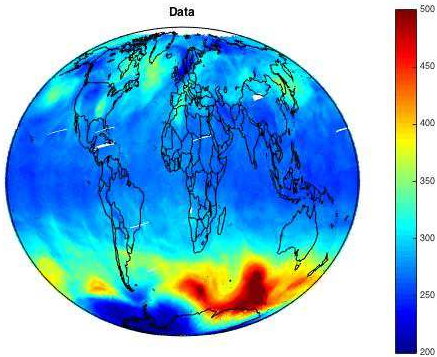}
		\end{tabular}
		\caption{Level 2 total column ozone data (in Dobson units) collected on October 1st, 1988, and analyzed by \citet{cressie2008fixed}.}\label{originalozone}
	\end{center}
\end{figure}

As an illustration, we analyze the ozone dataset used in \citet{cressie2008fixed}, which has become a benchmark dataset in the spatial statistics literature \citep{zhang2018smoothed}. This dataset consists of $\textit{n}=173,405$ values of total column ozone (TCO) in Dobson units (see Figure \eqref{originalozone}  for a plot of the data). The dataset was obtained through a Dobson spectrophotometer on board the Nimbus-7 polar orbiting satellite on October 1st, 1988. For details on how these data were collected see \citet{cressie2008fixed}. This dataset is made publically available by the Centre for Environmental Informatics at the University of Wollongong's National Institute for Applied Statistics Research Australia (\url{https://hpc.niasra.uow.edu.au/ckan/}).

\subsection{Analysis}
We present an analysis of the ozone dataset using ESD. We partition the data into a training set and a prediction set. We randomly generated three different 5\% missing at random  datasets for evaluating the prediction performance of all methods. A total of 5,000 MCMC iterations of the Gibbs sampler in Algorithm 1 were used. The first 1,000 iterations were treated as a burn-in. We informally check trace plots for convergence, and no lack of convergence was detected. Since \textit{d}=2, $\pmb{\psi}(\textbf{s})$ is an $r\times 2$ dimensional matrix, which we denote with $\pmb{\psi}(\textbf{s})=\left\lbrace\pmb{\phi}_{1}(\textbf{s}),\pmb{\phi}_{2}(\textbf{s})\right\rbrace$, where $\pmb{\phi}_{i}(\textbf{s})$ is an \textit{r}-dimensional vector, \textit{i}=1,2. Using the R-package \texttt{FRK}, we choose 92, 364 and 591 equally-spaced bisquare basis functions, which defines a 92, 364 and 591-dimensional vector $\pmb{\zeta}(\textbf{s})$ \citep{zammit2017frk}. Then, we set $\pmb{\phi}_{1}(\textbf{s})=\left\lbrace\textbf{0}_{3}^{\prime},\pmb{\zeta}(\textbf{s})^{\prime}\right\rbrace^{\prime}$, and we take $\pmb{\phi}_{2}(\textbf{s})=(1,\textbf{s}^{\prime},\textbf{0}_{r}^{\prime})^{\prime}$. This choice of $\pmb{\phi}_{2}(\textbf{s})$ isolates the effect of the latitude and longitude on the non-stationarity of the process. The covariates are defined to be $\textbf{X}(\textbf{s})=(1, \pmb{\zeta}(\textbf{s})^{\prime})^{\prime}$. 

\begin{figure}[H]
	\begin{center}
		\begin{tabular}{c}
			\includegraphics[width=.45\textwidth]{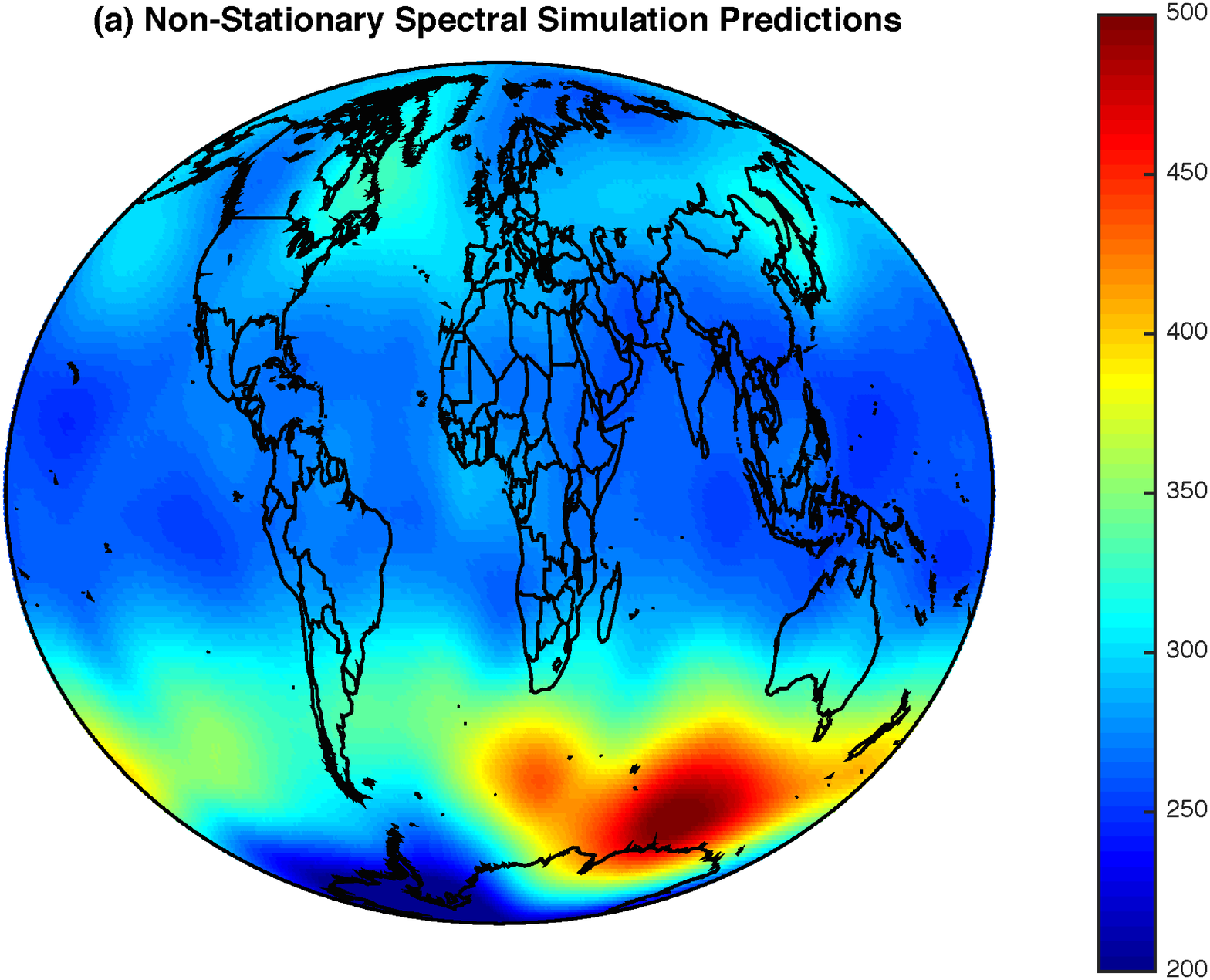}
			\includegraphics[width=.45\textwidth]{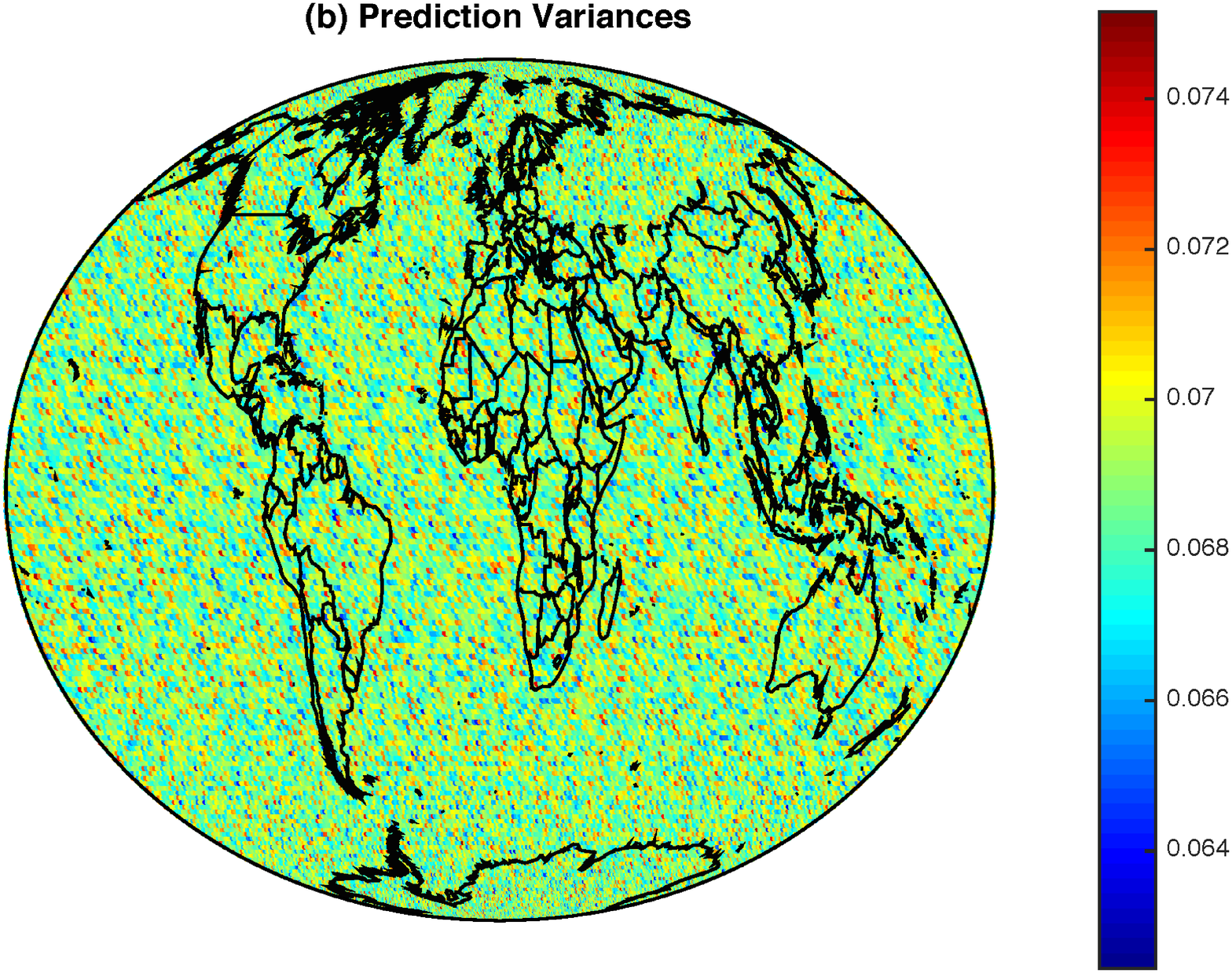}
		\end{tabular}
		\vspace{-50pt}
		\caption{Results for 92 basis function. In (a), we plot the posterior means (in Dobson units) from the model in (\ref{model}), which was implemented using the Gibbs sampler outlined in Algorithm 1. The corresponding posterior variances (in Dobson units squared) are presented in (b).} \label{r92}
	\end{center}
\end{figure}

\begin{figure}[H]
	\begin{center}
		\begin{tabular}{c}
			\includegraphics[width=.45\textwidth]{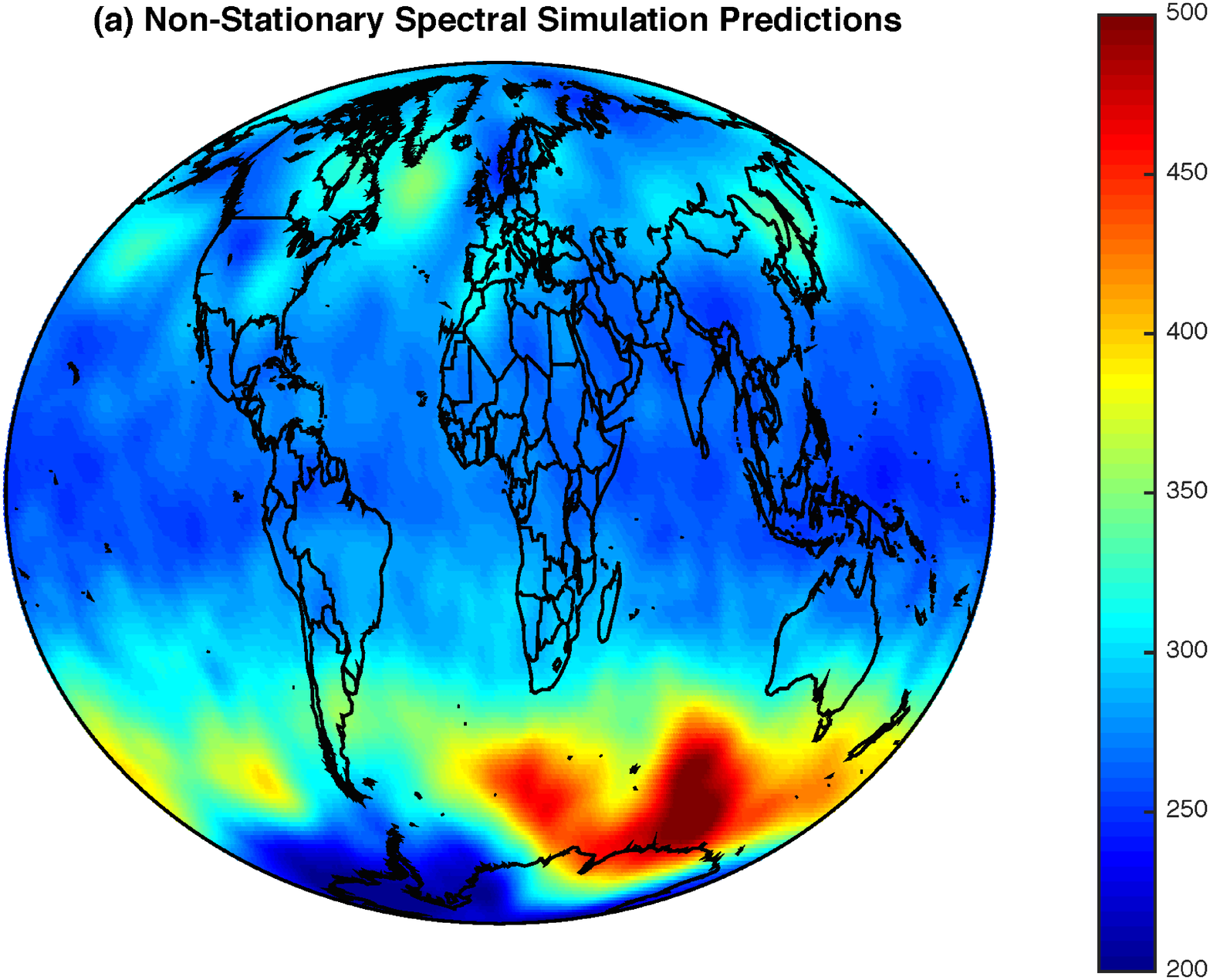}
			\includegraphics[width=.45\textwidth]{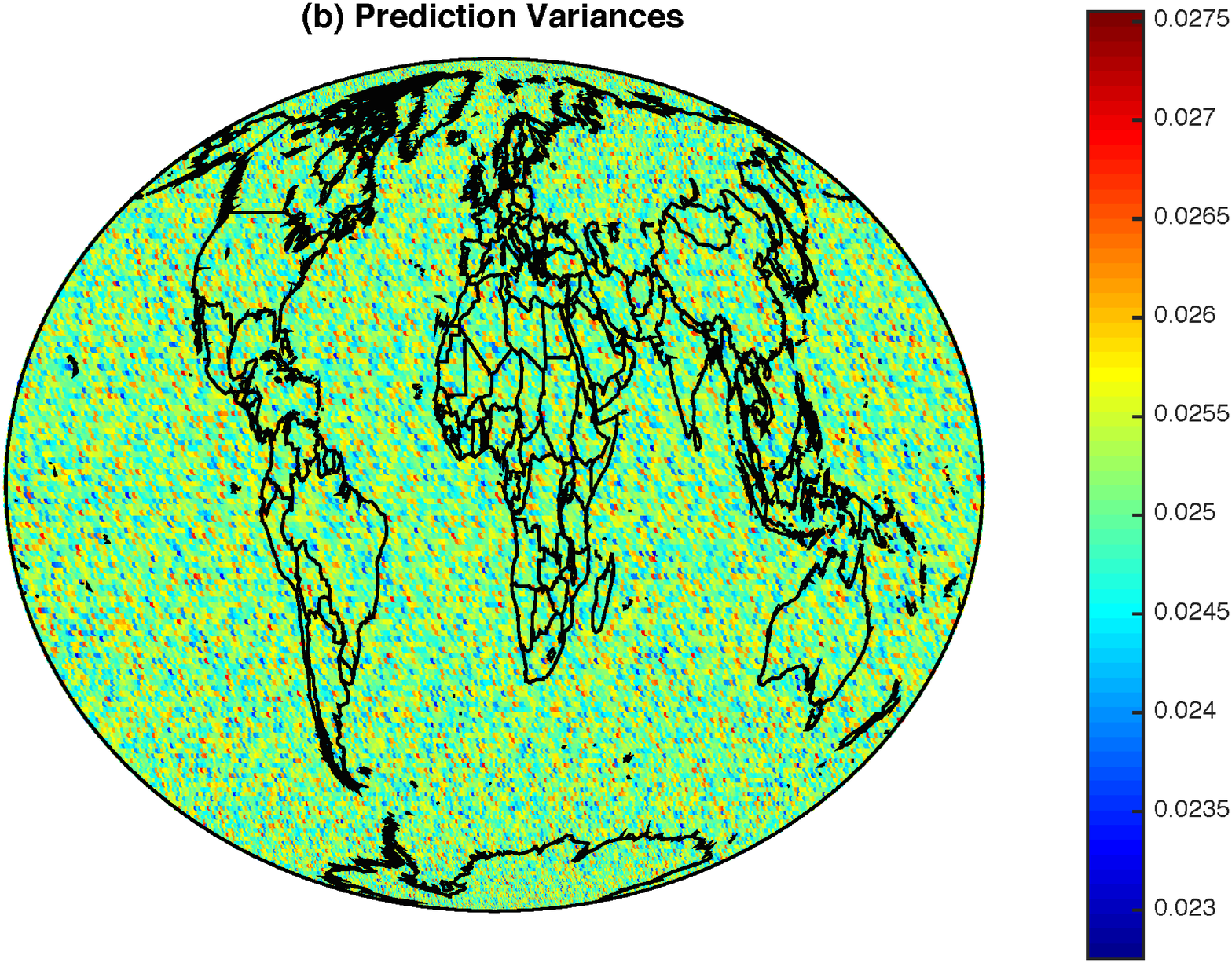}
		\end{tabular}
		\vspace{-50pt}
		\caption{Results for 364 basis function. In (a), we plot the posterior means (in Dobson units) from the model in (\ref{model}), which was implemented using the Gibbs sampler outlined in Algorithm 1. The corresponding posterior variances (in Dobson units squared) are presented in (b).} \label{r364}
	\end{center}
\end{figure}

\begin{figure}[H]
	\begin{center}
		\begin{tabular}{c}
			\includegraphics[width=.45\textwidth]{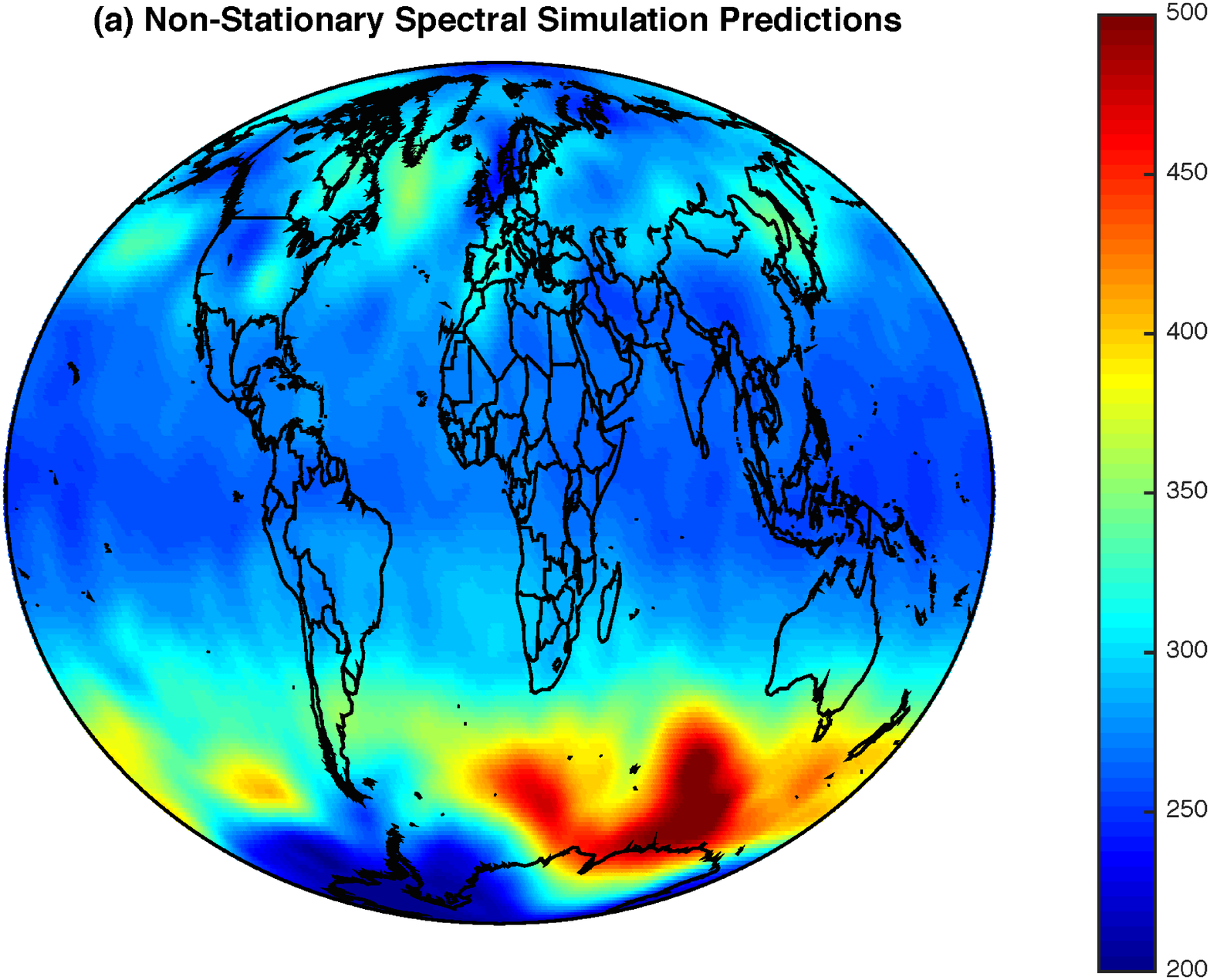}
			\includegraphics[width=.45\textwidth]{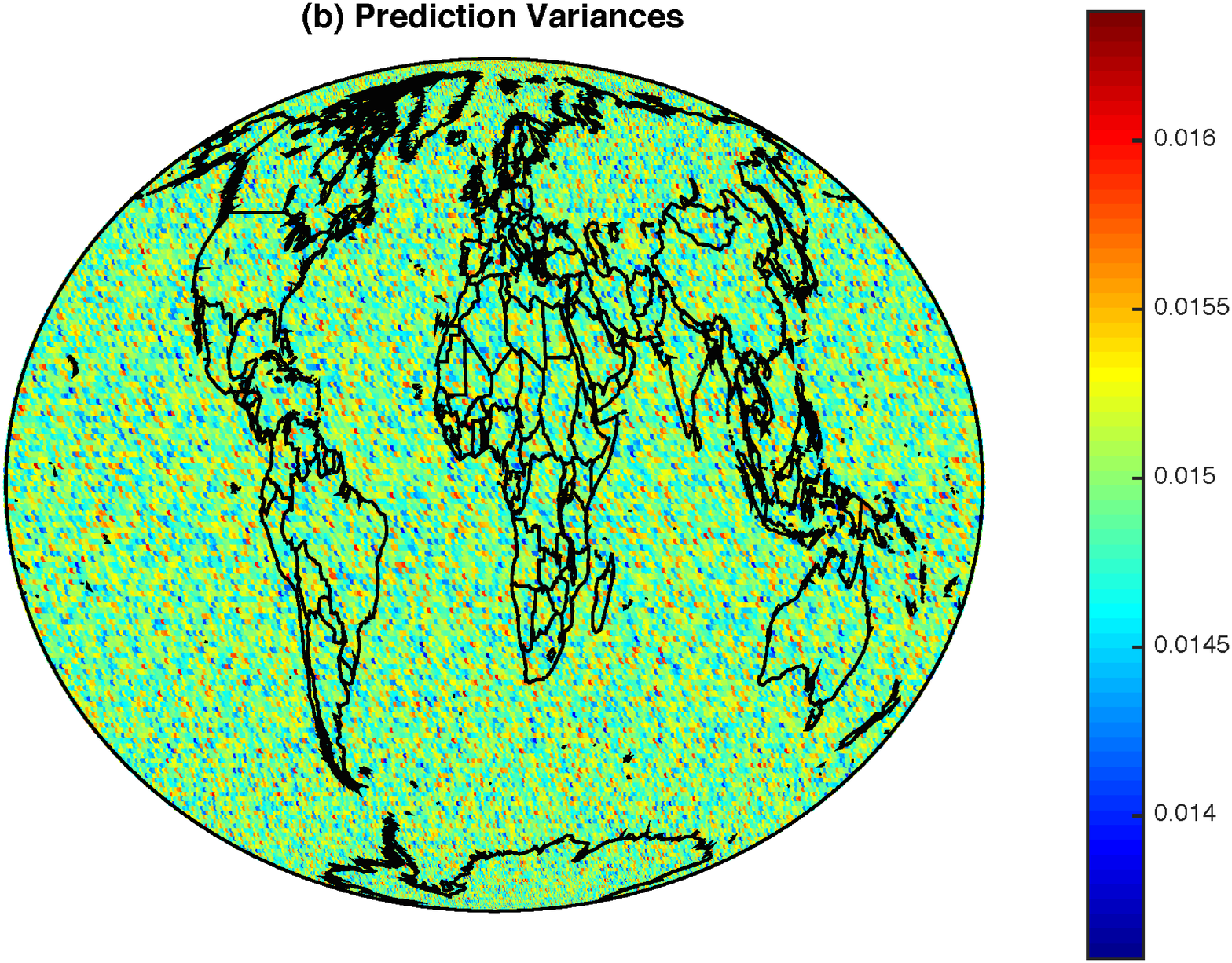}
		\end{tabular}
		\vspace{-50pt}
		\caption{Results for 591 basis function. In (a), we plot the posterior means (in Dobson units) from the model in (\ref{model}), which was implemented using the Gibbs sampler outlined in Algorithm 1. The corresponding posterior variances (in Dobson units squared) are presented in (b).} \label{r591}
	\end{center}
\end{figure}

Figure \eqref{r92} displays the prediction and prediction variances using non-stationary spectral simulation. Upon comparison of Figure \eqref{originalozone} to Figure (\ref{r92}a),Figure (\ref{r364}a) and Figure (\ref{r591}a) , we see that we obtain small in-sample error. Additionally, Figure (\ref{r92}b), Figure (\ref{r364}b) and Figure (\ref{r591}b) shows that our prediction error is relatively constant over the globe. We randomly select 5\% of the total observations to act as a validation dataset, and compute the root mean squared prediction error (RMSPE). Specifically, we compute the average square distance between the validation data and its corresponding prediction, and then we take the square root. RMSPE for our method is around 16.55, 9.86 and 7.52 for 92, 364 and 591 basis functions, respectively (see Table \ref{table2}). We also computed the RMSPE for the fixed rank kriging method as implemented through the R-package \texttt{FRK} \citep{zammit2017frk}. The FRK predictor is based on $\textbf{X}(\textbf{s})=1$ and uses $\pmb{\zeta}(\textbf{s})$ as its basis set. The RMSPE for FRK is approximately 76.41, and hence, we outperform the FRK predictor in terms of RMSPE. We do compare to other methods as well. In \citet{zhang2018smoothed} paper, they compare the Smoothed Full-Scale Approximation (SFSA), Full-Scale Approximation using a block modulating function (FSAB) \citep{sang2012full}, NNGP, and a local Gaussian process method with adaptive local designs (LaGP) \citep{gramacy2015local}. Their results show that the RMSPE for SFSA, NNGP and FSAB are all around 27, and the RMSPE for LaGP is around 38. Thus, our method also outperforms these methods in terms of RMSPE. However, the general Vecchia approximation has a slightly better result. With RMSPE equal to roughly 5, which is smaller than our RMSPE of 7.52. This consistent with our simulation results that showed that in smoother nonstationary settings the general Vecchia approximation performs similar or better than ESD. 



We also include the computation times in Table \eqref{Time}. Our method is less competitive. Although we avoid storing and inverting a high dimensional covariance matrix, we require nested loops, which can be computationally intensive (i.e., a loop in the Gibbs sampler and a loop over $i=1\ldots K$ in Theorem 2). However, we are able to produce spatial predictions.

\begin{table}[H]
	\begin{center}
		\caption{Sensitivity to the number of basis functions}
		\label{table2}
		\begin{tabular}{c|c} 
			\textbf{Total size of basis function} & \textbf{RMSPE}\\
			\hline
			92 &  16.55\\
			\hline
			364&   9.86\\
			\hline
			591&    7.52\\
			\hline
		\end{tabular}
	\end{center}
\end{table}

\begin{table}[H]
	\begin{center}
		\caption{Computation Time by Basis Function.}
		\label{Time}
		\begin{tabular}{c|c|c} 
			\textbf{Method} &\textbf{Number of basis function}& \textbf{Time (seconds)}\\
			\hline
			ESD &92  &13925\\
			 &364  &111685\\
			 &591  &273024\\
			\hline
			Vecchia Approximation&- &200  \\
			\hline
		\end{tabular}
	\end{center}
\end{table}

\begin{figure}[t!]
	\begin{center}
		\vspace{-30pt}
		\begin{tabular}{c}
			\includegraphics[width=\textwidth]{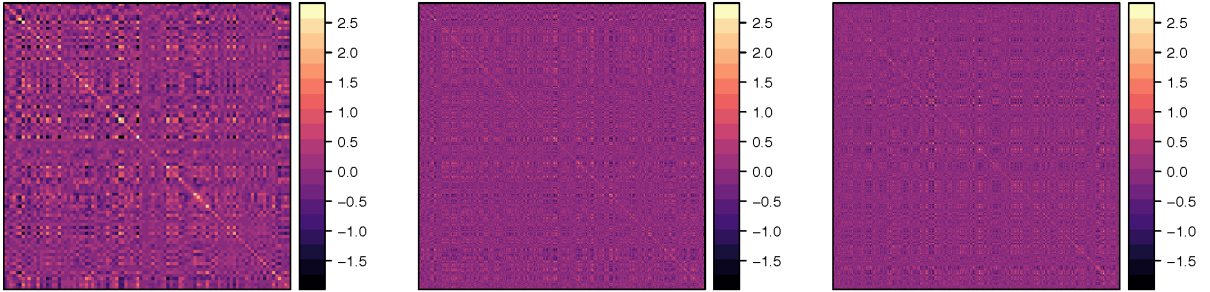}
		\end{tabular}
		\vspace{-100pt}
		\caption{Posterior Covariance Matrix Image Plot. From left to right is 92, 364, 591 basis function.} \label{PosVar}
	\end{center}
\end{figure}

In terms of inference on parameters, we are particularly interested in $\pmb{\eta}$. This is because when $\pmb{\eta}$ is zero we obtain a stationary process (see Theorem 1). In Figure \eqref{PosVar} we plot the posterior covariance matrix. As the $r$ increases, we see the variances and covariances appear to be close to zero suggesting that $\pmb{\eta}$ is close to zero. This suggests that the process is smooth, which is consistent with our previous results. However, several credible intervals for elements of $\pmb{\eta}$ do not contain zero, which suggests that nonstationarity is present in this dataset.

\section{Discussion}
Bayesian analysis of big Gaussian spatial data is a challenging and important problem. We propose a Bayesian approach using non-stationary spectral simulation. To develop non-stationary spectral simulation we combine Bochner's theorem with dimension expansion \citep{perrin2003nonstationarity}, and apply \citet{mejia1974synthesis}'s spectral simulation method. The advantage is that no large matrix inversion or storage is needed to approximately simulate a non-stationary full-rank Gaussian process. Additionally, the proposed method is extremely broad, since every positive definite non-stationary covariance function can be written according to \eqref{thm1}.

In Section 4, the simulation study is used to show a scenario where our approach outperforms the nearest neighbor Gaussian process \citep[NNGP;][]{datta2016hierarchical} model and Vecchia approximation \citep{katzfuss2017general}. We generate data that is different from our model, and we find our method has better result in different scenarios based on how nonlinear the process is. In Section 5, we analyze the total column ozone dataset from \citet{cressie2008fixed}. We obtain predictions that have small in-sample error, and outperforms fixed rank kriging (FRK), SFSA, FSAB, NNGP, and LaGP in terms of out-of-sample error. The Vecchia approximation has a slightly better RMSPE. Additionally, our framework allows one to perform inference on the presence of nonstationarity.

Environmental studies are often based on high-dimensional spatial Gaussian datasets with complex patterns of non-stationarity. Several studies focus on simplifying matrix valued operations and storage \citep{higdon1999non,paciorek2006spatial,cressie2008fixed,banerjee2008gaussian,lindgren2011explicit,nychka2015multiresolution}. Thus, our ``matrix free" approach offers a unique solution to this important problem. 

\section*{Acknowledgments} Jonathan Bradley's research was partially supported by the U.S. National Science Foundation (NSF) grant SES-1853099.

\section*{Appendix A: Proofs}
\noindent
\textbf{Proof of Theorem 1:}\\
It follows from \citet{perrin2003nonstationarity} that for every non-stationary positive definite function $\textit{C}$ and every pair of locations $s_{1}$ and $s_{2}$ there exists a $\textbf{w}_{1}$ and $\textbf{w}_{2}$ such that $C(\textbf{s}_{1},\textbf{s}_{2})=\rho\left\lbrace\bigl(\begin{smallmatrix}
\textbf{s}_{1}\\ \textbf{w}_{1}
\end{smallmatrix}\bigr),\bigl(\begin{smallmatrix}
\textbf{s}_{2}\\ \textbf{w}_{2}
\end{smallmatrix}\bigr)\right\rbrace$, where $\rho$ is a stationary covariogram. Let $f$ be the function that maps a generic location $\textbf{s}\in D$ to it's corresponding expanded dimension $\textbf{w}\in\mathbb{R}^{d}$.

It follows from Bochner's theorem \citep{bochner1959lectures} that $\rho\left\lbrace\bigl(\begin{smallmatrix}
\textbf{s}_{i}\\ \textbf{\textit{f}}(\textbf{s}_{i})
\end{smallmatrix}\bigr),\bigl(\begin{smallmatrix}
\textbf{s}_{j}\\ \textbf{\textit{f}}(\textbf{s}_{j})
\end{smallmatrix}\bigr)\right\rbrace$ is positive definite (and equivalently so is $C(s_{i},s_{j})$) if and only if
\begin{align*}
	&C(s_{i},s_{j})=\rho\left\lbrace\bigl(\begin{smallmatrix}
	\textbf{s}_{i}\\ \textbf{\textit{f}}(\textbf{s}_{i})
	\end{smallmatrix}\bigr),\bigl(\begin{smallmatrix}
	\textbf{s}_{j}\\ \textbf{\textit{f}}(\textbf{s}_{j})
	\end{smallmatrix}\bigr)\right\rbrace=\int_{-\infty}^{\infty}\cos\left\lbrace\{\textbf{\textit{f}}(\textbf{s}_{i})-\textbf{\textit{f}}(\textbf{s}_{j})\}^{\prime}\pmb{\omega}_{1}+(\textbf{s}_{i}-\textbf{s}_{j})^{\prime}\pmb{\omega}_{2}\right\rbrace G(d\omega).
\end{align*}
This completes the result.\\
\textbf{Proof of Theorem 2:}\\
We have that,
\begin{equation*}
E\{\hat{\pmb{\nu}}(\textbf{s})\}=\frac{(2)^{\frac{1}{2}}}{2\pi\sigma_{\nu}}\int_{-\infty}^{\infty}\int_{-\pi}^{\pi}\cos(\textbf{\textit{f}}(\textbf{s})^{\prime}\pmb{\omega}_{1}+\textbf{s}^{\prime}\bm{\omega}_{2}+\kappa)g(\pmb{\omega})d\kappa d\pmb{\omega}=0,
\end{equation*}
since $\int_{-\pi}^{\pi}\cos(\kappa)d\kappa=\int_{-\pi}^{\pi}\sin(\kappa)d\kappa=0$. Also,
\begin{align*}
&E\{\hat{\pmb{\nu}}(s_{i})\hat{\pmb{\nu}}(s_{j})\}\\
&=2\sigma^{2}_{\nu}E[\cos(\textbf{\textit{f}}(\textbf{s}_{i})^{\prime}\pmb{\omega}_{1}+\textbf{s}_{i}^{\prime}\pmb{\omega}_{2}+\kappa_{1})\cos(\textbf{\textit{f}}(\textbf{s}_{j})^{\prime}\pmb{\omega}_{1}+\textbf{s}_{j}^{\prime}\pmb{\omega}_{2}+\kappa_{1})]\\
&=\int\cos\{(\textbf{\textit{f}}(\textbf{s}_{i})^{\prime}-\textbf{\textit{f}}(\textbf{s}_{j})^{\prime})\pmb{\omega}_{1}+(s_{i}-s_{j})\pmb{\omega}_{2}\}g(\pmb{\omega}_{1})g(\pmb{\omega}_{2})d\pmb{\omega}_{1}d\pmb{\omega}_{2}\\
&=\rho\left\lbrace\bigl(\begin{smallmatrix}
\textbf{s}_{i}\\ \textbf{f}(\textbf{s}_{i})
\end{smallmatrix}\bigr)-\bigl(\begin{smallmatrix}
\textbf{s}_{j}\\ \textbf{f}(\textbf{s}_{j})	
\end{smallmatrix}\bigr)\right\rbrace=C(s_{i},s_{j}).
\end{align*}
since $\rho(\cdot)$ is a stationary covariogram it follow from \citet[][pg. 204]{cressie1993statistics} that $\widetilde{\pmb{\nu}}$ converges to a Gaussian process.

\section*{Appendix B: Full Conditional Distributions}
In this section, we derive the full conditional distribution of our parameters and random effects, which we use within the Gibbs sampler outlined in Algorithm 1.
\begin{enumerate}
	\item The full conditional distribution for $\pmb{\beta}$ is
	\begin{equation*}
	f(\pmb{\beta}\vert\cdot)\propto\exp\left\lbrace-\frac{(\textbf{Z}-\textbf{X}\pmb{\beta}-\textbf{O}\pmb{\nu})^{\prime}\textbf{V}_{\epsilon}^{-1}(\textbf{Z}-\textbf{X}\pmb{\beta}-\textbf{O}\pmb{\nu})}{2}\right\rbrace\exp\left\lbrace-\frac{\pmb{\beta}^{\prime}\pmb{\beta}}{2\sigma_{\beta}^{2}}\right\rbrace
	\end{equation*}
	\begin{equation*}
	\propto\exp\left\lbrace-\frac{\pmb{\beta}^{\prime}(\textbf{X}^{\prime}\textbf{V}_{\epsilon}^{-1}\textbf{X}+\sigma_{\beta}^{-2}\textbf{I}_{p})\pmb{\beta}}{2}+\pmb{\beta}^{\prime}\textbf{X}^{\prime}\textbf{V}_{\epsilon}^{-1}(\textbf{Z}-\textbf{O}\pmb{\nu})\right\rbrace
	\end{equation*}
	\begin{equation*}
	\propto\exp\left\lbrace-\frac{\pmb{\beta}^{\prime}\pmb{\Sigma}_{*}^{-1}\pmb{\beta}}{2}+\pmb{\beta}^{\prime}\pmb{\Sigma}_{*}^{-1}\pmb{\mu}_{*}\right\rbrace
	\end{equation*}
	\begin{equation*}
	\propto \exp\left\lbrace-\frac{1}{2}(\pmb{\beta}-\pmb{\mu}_{*})^{\prime}\pmb{\Sigma}_{*}^{-1}(\pmb{\beta}-\pmb{\mu}_{*})\right\rbrace.
	\end{equation*}
	Thus, $\pmb{\beta}\sim \text{N}(\pmb{\mu}_{*},\pmb{\Sigma}_{*})$, where $\pmb{\mu}_{*}=\pmb{\Sigma}_{*}^{-1}\pmb{X}'\textbf{V}_{\epsilon}^{-1}(\textbf{Z}-\textbf{O}\pmb{\nu})$ and $\pmb{\Sigma}_{*}^{-1}=\pmb{X}'\textbf{V}_{\epsilon}^{-1}\textbf{X}+\sigma_{\beta}^{-2}\textbf{I}_{p}$, $\textbf{X}=(\textbf{X}(\textbf{s}_{1})\ldots \textbf{X}(\textbf{s}_{n}))^{\prime}$.
	
	\item First, we sample $\widetilde{\pmb{\nu}}$ using non-stationary spectral simulation. Then the full-conditional distribution is
	\begin{equation*}
	f(\pmb{\nu}\vert\cdot)\propto\exp\left\lbrace-\frac{(\textbf{Z}-\textbf{X}\pmb{\beta}-\textbf{O}\pmb{\nu})^{\prime}\textbf{V}_{\epsilon}^{-1}(\textbf{Z}-\textbf{X}\pmb{\beta}-\textbf{O}\pmb{\nu})}{2}\right\rbrace f(\pmb{\nu}\vert\widetilde{\pmb{\nu}},\delta,\pmb{\theta}),
	\end{equation*}
	where $f(\pmb{\nu}\vert\widetilde{\pmb{\nu}}, \delta,\pmb{\theta})=\exp\left\lbrace-\frac{(\pmb{\nu}-\widetilde{\pmb{\nu}})^{'}(\pmb{\nu}-\widetilde{\pmb{\nu}})}{2\delta^{2}}\right\rbrace$. Then,
	\begin{equation*}
	f(\pmb{\nu}\vert\cdot)\propto\exp\left\lbrace-\frac{\pmb{\nu}^{\prime}(\delta^{-2}\textbf{I}+\textbf{O}'\textbf{V}_{\epsilon}^{-1}\textbf{O})\pmb{\nu}}{2}+\nu^{\prime}\left(\textbf{O}'\textbf{V}_{\epsilon}^{-1}(\textbf{Z}-\textbf{X}\pmb{\beta})+\frac{1}{\delta^{2}}\widetilde{\pmb{\nu}}\right)\right\rbrace.
	\end{equation*}
    This gives
	\begin{equation*}
	f(\pmb{\nu}\vert\cdot)\propto \text{N}(\pmb{\mu}^{*},\pmb{\Sigma}^{*}),
	\end{equation*}
	where	$\pmb{\mu}^{*}=\pmb{\Sigma}^{*}\{\textbf{O}'\textbf{V}_{\epsilon}^{-1}(\textbf{Z}-\textbf{X}\pmb{\beta})+\frac{1}{\delta^{2}}\widetilde{\pmb{\nu}}\}$, and $(\pmb{\Sigma}^{*})^{-1}=\delta^{-2}\textbf{I}+\textbf{O}'\textbf{V}_{\epsilon}^{-1}\textbf{O}$.
	\item The approximated full conditional distribution for $\pmb{\eta}$ is given by
	\begin{equation*}
	f(\pmb{\eta}\vert\cdot) \propto \exp\left\lbrace-\frac{\pmb{\eta}^{\prime}\pmb{\eta}}{2\sigma_{\eta}^{2}}\right\rbrace\exp(-\frac{(\pmb{\nu}-\widetilde{\pmb{\nu}})^{\prime}(\pmb{\nu}-\widetilde{\pmb{\nu}})}{2\delta^{2}}),
	\end{equation*}
	where recall $\widetilde{\pmb{\nu}}$ is a function of $\pmb{\eta}$. We use Metropolis-Hasting to sample $\pmb{\eta}$ and we use a multivariate normal distribution for the proposal distribution.
	
	
	\item For $\sigma^2_{\nu}$, we use a Inverse Gamma prior distribution,
	\begin{equation*}
	f(\sigma_{\nu}^{2}\vert\cdot) \propto (\sigma_{\nu}^{2})^{-(\alpha_{1}+\frac{n}{2})-1} \exp\left\lbrace{-\frac{\pmb{\nu}'_{m}C(\pmb{\theta})^{-1}\pmb{\nu}_{m}}{2\sigma_{\nu}^{2}}+\sigma_{\nu}^{-2}\beta_{1}}\right\rbrace,
	\end{equation*}
	which is an inverse gamma distribution with shape parameter $\alpha_{1}+\frac{n}{2}$ and scale parameter $\frac{\pmb{\nu}'_{m}C(\pmb{\theta})^{-1}\pmb{\nu}_{m}}{2}+\beta_{1}$.
	
	\item The full conditional distribution of $\sigma_{\beta}^{2}$ is easily obtained and given by,
	\begin{equation*}
	f(\sigma_{\beta}^{2}\vert\cdot) \propto (\sigma_{\beta}^{2})^{-(\alpha_{2}+\frac{p}{2})-1} \exp\left\lbrace{-\frac{\pmb{\beta}^{\prime}\pmb{\beta}}{2\sigma_{\beta}^{2}}+\sigma_{\beta}^{-2}\beta_{2}}\right\rbrace,
	\end{equation*}
	which is an inverse gamma distribution with shape parameter $\alpha_{2}+\frac{p}{2}$ and scale parameter $\frac{\pmb{\beta}^{\prime}\pmb{\beta}}{2}+\beta_{2}$.
	
	\item The full conditional distribution of $\sigma_{\eta}^{2}$ is easily obtained and given by,
	\begin{equation*}
	f(\sigma_{\eta}^{2}\vert\cdot) \propto (\sigma_{\eta}^{2})^{-(\alpha_{3}+\frac{r}{2})-1} \exp\left\lbrace{-\frac{\pmb{\eta}^{\prime}\pmb{\eta}}{2\sigma_{\eta}^{2}}+\sigma_{\eta}^{-2}\beta_{3}}\right\rbrace,
	\end{equation*}
	which is an inverse gamma distribution with shape parameter $\alpha_{3}+\frac{r}{2}$ and scale parameter $\frac{\pmb{\eta}^{\prime}\pmb{\eta}}{2}+\beta_{3}$.
	
	\item The prior for $\phi$ is $\text{Uniform Distribution}(L,U)$, and the full conditional distribution for $\phi$ follows that,
	\begin{equation*}
	f(\phi\vert\cdot)\propto f(\pmb{\nu}_{m}\mid\widetilde{\pmb{\nu}}, \sigma_{\nu}^2,\phi)I_{(L\le\phi\le U)}
	\end{equation*}
	
	\item The full conditional distribution of $\delta^2$ is easily obtained and given by,
	\begin{equation*}
	f(\delta^2\vert\cdot)\propto(\delta^{2})^{-(\alpha_{4}+\frac{n}{2})-1}\exp\left\lbrace-\frac{(\pmb{\nu}-\widetilde{\pmb{\nu}})^{\prime}(\pmb{\nu}-\widetilde{\pmb{\nu}})}{2\delta^2}+\frac{\beta_{4}}{\delta^2}\right\rbrace.\label{delta}
	\end{equation*}
	which is an inverse gamma distribution with shape parameter $\alpha_{4}+\frac{n}{2}$ and scale parameter
	 $\frac{(\pmb{\nu}-\widetilde{\pmb{\nu}})'(\pmb{\nu}-\widetilde{\pmb{\nu}})}{2}+\beta_{4}$.
	 
	 \item The full conditional distribution of $\sigma_{\epsilon}^2$ is easily obtained and given by,
	 \begin{equation*}
	 f(\sigma_{\epsilon}^2\vert\cdot)\propto(\sigma_{\epsilon}^2)^{-(\alpha_{5}+\frac{n}{2})-1}\exp\left\lbrace-\frac{(\textbf{Z}-\textbf{X}\pmb{\beta}-\pmb{\nu})^{\prime}(\textbf{Z}-\textbf{X}\pmb{\beta}-\pmb{\nu})}{2\sigma_{\epsilon}^2}+\frac{\beta_{5}}{\sigma_{\epsilon}^2}\right\rbrace.\label{tau}
	 \end{equation*}
	 which is an inverse gamma distribution with shape parameter $\alpha_{5}+\frac{n}{2}$ and scale parameter
	 $\frac{(\textbf{Z}-\textbf{X}\pmb{\beta}-\pmb{\nu})^{\prime}(\textbf{Z}-\textbf{X}\pmb{\beta}-\pmb{\nu})}{2}+\beta_{5}$.

\end{enumerate}
\bibliographystyle{jasa}
\bibliography{main}

\begin{thebibliography}{74}
\newcommand{\enquote}[1]{``#1''}
\expandafter\ifx\csname natexlab\endcsname\relax\def\natexlab#1{#1}\fi

\bibitem[\protect\citename{Anderes and Stein, }2011]{anderes2011local}
Anderes, E.~B. and Stein, M.~L. (2011).
\newblock \enquote{Local likelihood estimation for nonstationary random
  fields.}
\newblock {\em Journal of Multivariate Analysis\/}, 102, 3, 506--520.

\bibitem[\protect\citename{AO~Finley, }2017]{spNNGP}
AO~Finley, A~Datta, S.~B. (2017).
\newblock \enquote{Package `spNNGP'.}
\newblock https://cran.r-project.org/web/packages/spNNGP/index.html.

\bibitem[\protect\citename{Banerjee et~al., }2015]{banerjee2014hierarchical}
Banerjee, S., Carlin, B.~P., and Gelfand, A.~E. (2015).
\newblock {\em Hierarchical modeling and analysis for spatial data\/}.
\newblock Boca Raton, FL, CRC Press.

\bibitem[\protect\citename{Banerjee et~al., }2008]{banerjee2008gaussian}
Banerjee, S., Gelfand, A.~E., Finley, A.~O., and Sang, H. (2008).
\newblock \enquote{Gaussian predictive process models for large spatial data
  sets.}
\newblock {\em Journal of the Royal Statistical Society: Series B (Statistical
  Methodology)\/}, 70, 4, 825--848.

\bibitem[\protect\citename{Bochner, }1959]{bochner1959lectures}
Bochner, S. (1959).
\newblock {\em Lectures on Fourier Integrals\/}.
\newblock No.~42. Princeton University Press.

\bibitem[\protect\citename{Bornn et~al., }2017]{bornn2017use}
Bornn, L., Pillai, N.~S., Smith, A., and Woodard, D. (2017).
\newblock \enquote{The use of a single pseudo-sample in approximate Bayesian
  computation.}
\newblock {\em Statistics and Computing\/}, 27, 3, 583--590.

\bibitem[\protect\citename{Bornn et~al., }2012]{bornn2012modeling}
Bornn, L., Shaddick, G., and Zidek, J.~V. (2012).
\newblock \enquote{Modeling nonstationary processes through dimension
  expansion.}
\newblock {\em Journal of the American Statistical Association\/}, 107, 497,
  281--289.

\bibitem[\protect\citename{Bradley et~al., }2011]{bradley2011}
Bradley, J.~R., Cressie, N., and Shi, T. (2011).
\newblock \enquote{Selection of rank and basis functions in the Spatial Random
  Effects model.}
\newblock In {\em Proceedings of the 2011 Joint Statistical Meetings\/},
  3393--3406. Alexandria, VA: American Statistical Association.

\bibitem[\protect\citename{Bradley et~al., }2016]{bradley2016comparison}
Bradley, J.~R., Cressie, N., Shi, T., et~al. (2016).
\newblock \enquote{A comparison of spatial predictors when datasets could be
  very large.}
\newblock {\em Statistics Surveys\/}, 10, 100--131.

\bibitem[\protect\citename{Bradley et~al., }2018]{bradleyhierarchical}
Bradley, J.~R., Wikle, C.~K., and Holan, S.~H. (2018).
\newblock \enquote{Hierarchical Models for Spatial Data with Errors that are
  Correlated with the Latent Process.}

\bibitem[\protect\citename{Castruccio and Guinness,
  }2017]{castruccio2017evolutionary}
Castruccio, S. and Guinness, J. (2017).
\newblock \enquote{An evolutionary spectrum approach to incorporate large-scale
  geographical descriptors on global processes.}
\newblock {\em Journal of the Royal Statistical Society: Series C (Applied
  Statistics)\/}, 66, 2, 329--344.

\bibitem[\protect\citename{Cressie and Johannesson, }2008]{cressie2008fixed}
Cressie, N. and Johannesson, G. (2008).
\newblock \enquote{Fixed rank kriging for very large spatial data sets.}
\newblock {\em Journal of the Royal Statistical Society: Series B (Statistical
  Methodology)\/}, 70, 1, 209--226.

\bibitem[\protect\citename{Cressie, }1993]{cressie1993statistics}
Cressie, N.~A. (1993).
\newblock \enquote{\textit{Statistics for spatial data}.}

\bibitem[\protect\citename{Cressie and Wikle, }2011]{cressie2011statistics}
Cressie, N.~A. and Wikle, C. (2011).
\newblock \enquote{Statistics for Spatio-Temporal Data.}

\bibitem[\protect\citename{Datta et~al.,
  }2016{\natexlab{a}}]{datta2016hierarchical}
Datta, A., Banerjee, S., Finley, A.~O., and Gelfand, A.~E.
  (2016{\natexlab{a}}).
\newblock \enquote{Hierarchical nearest-neighbor Gaussian process models for
  large geostatistical datasets.}
\newblock {\em Journal of the American Statistical Association\/}, 111, 514,
  800--812.

\bibitem[\protect\citename{Datta et~al., }2016{\natexlab{b}}]{datta2016nearest}
--- (2016{\natexlab{b}}).
\newblock \enquote{On nearest-neighbor Gaussian process models for massive
  spatial data.}
\newblock {\em Wiley Interdisciplinary Reviews: Computational Statistics\/}, 8,
  5, 162--171.

\bibitem[\protect\citename{Finley et~al., }2013]{finley2013spbayes}
Finley, A.~O., Banerjee, S., and Gelfand, A.~E. (2013).
\newblock \enquote{spBayes for large univariate and multivariate
  point-referenced spatio-temporal data models.}
\newblock {\em \textit{Journal of Statistical Software}\/}.

\bibitem[\protect\citename{Friedman et~al., }1991]{friedman1991multivariate}
Friedman, J.~H. et~al. (1991).
\newblock \enquote{Multivariate adaptive regression splines.}
\newblock {\em The annals of statistics\/}, 19, 1, 1--67.

\bibitem[\protect\citename{Fuentes, }2002]{fuentes2002spectral}
Fuentes, M. (2002).
\newblock \enquote{Spectral methods for nonstationary spatial processes.}
\newblock {\em Biometrika\/}, 89, 1, 197--210.

\bibitem[\protect\citename{Fuentes et~al., }2008]{fuentes2008class}
Fuentes, M., Chen, L., and Davis, J.~M. (2008).
\newblock \enquote{A class of nonseparable and nonstationary spatial temporal
  covariance functions.}
\newblock {\em Environmetrics\/}, 19, 5, 487--507.

\bibitem[\protect\citename{Fuentes and Smith, }2001]{fuentes2001new}
Fuentes, M. and Smith, R.~L. (2001).
\newblock \enquote{A new class of nonstationary spatial models.}
\newblock Tech. rep., unpublished manuscript, available at
  www.stat.ncsu.edu/information/library/papers/nonstat.ps.

\bibitem[\protect\citename{Fuglstad et~al., }2015]{fuglstad2015interpretable}
Fuglstad, G.-A., Simpson, D., Lindgren, F., and Rue, H. (2015).
\newblock \enquote{Interpretable priors for hyperparameters for Gaussian random
  fields.}
\newblock {\em arXiv preprint arXiv:1503.00256\/}.

\bibitem[\protect\citename{Gelfand, }2000]{gelfand2000gibbs}
Gelfand, A.~E. (2000).
\newblock \enquote{Gibbs sampling.}
\newblock {\em Journal of the American statistical Association\/}, 95, 452,
  1300--1304.

\bibitem[\protect\citename{Gelfand and Smith, }1990]{gelfand1990sampling}
Gelfand, A.~E. and Smith, A.~F. (1990).
\newblock \enquote{Sampling-based approaches to calculating marginal
  densities.}
\newblock {\em Journal of the American statistical association\/}, 85, 410,
  398--409.

\bibitem[\protect\citename{Gelman, }2006]{gelman2006prior}
Gelman, A. (2006).
\newblock \enquote{Prior distributions for variance parameters in hierarchical
  models (comment on article by Browne and Draper).}
\newblock {\em Bayesian analysis\/}, 1, 3, 515--534.

\bibitem[\protect\citename{Gramacy and Apley, }2015]{gramacy2015local}
Gramacy, R.~B. and Apley, D.~W. (2015).
\newblock \enquote{Local Gaussian process approximation for large computer
  experiments.}
\newblock {\em Journal of Computational and Graphical Statistics\/}, 24, 2,
  561--578.

\bibitem[\protect\citename{Guinness and Katzfuss, }2018]{guinness2018gpgp}
Guinness, J. and Katzfuss, M. (2018).
\newblock \enquote{GpGp: Fast Gaussian Process Computation Using Vecchia’s
  Approximation.}
\newblock {\em R package version 0.1. 0\/}.

\bibitem[\protect\citename{Guinness and Stein,
  }2013]{guinness2013interpolation}
Guinness, J. and Stein, M.~L. (2013).
\newblock \enquote{Interpolation of nonstationary high frequency
  spatial--temporal temperature data.}
\newblock {\em The Annals of Applied Statistics\/}, 7, 3, 1684--1708.

\bibitem[\protect\citename{Hastings, }1970]{hastings1970monte}
Hastings, W.~K. (1970).
\newblock \enquote{Monte Carlo sampling methods using Markov chains and their
  applications.}
\newblock {\em Biometrika\/}, 57, 1, 97--109.

\bibitem[\protect\citename{Heaton et~al., }2017]{heaton2017methods}
Heaton, M.~J., Datta, A., Finley, A., Furrer, R., Guhaniyogi, R., Gerber, F.,
  Gramacy, R.~B., Hammerling, D., Katzfuss, M., Lindgren, F., et~al. (2017).
\newblock \enquote{Methods for Analyzing Large Spatial Data: A Review and
  Comparison.}
\newblock {\em arXiv preprint arXiv:1710.05013\/}.

\bibitem[\protect\citename{Higdon, }1998]{higdon1998process}
Higdon, D. (1998).
\newblock \enquote{A process-convolution approach to modelling temperatures in
  the North Atlantic Ocean.}
\newblock {\em Environmental and Ecological Statistics\/}, 5, 2, 173--190.

\bibitem[\protect\citename{Higdon et~al., }1999]{higdon1999non}
Higdon, D., Swall, J., and Kern, J. (1999).
\newblock \enquote{Non-stationary spatial modeling.}
\newblock {\em Bayesian statistics\/}, 6, 1, 761--768.

\bibitem[\protect\citename{Horrell and Stein, }2017]{horrell2017half}
Horrell, M.~T. and Stein, M.~L. (2017).
\newblock \enquote{Half-spectral space--time covariance models.}
\newblock {\em Spatial Statistics\/}, 19, 90--100.

\bibitem[\protect\citename{Im et~al., }2007]{im2007semiparametric}
Im, H.~K., Stein, M.~L., and Zhu, Z. (2007).
\newblock \enquote{Semiparametric estimation of spectral density with irregular
  observations.}
\newblock {\em Journal of the American Statistical Association\/}, 102, 478,
  726--735.

\bibitem[\protect\citename{Kalkhan, }2011]{kalkhan2011spatial}
Kalkhan, M.~A. (2011).
\newblock {\em Spatial statistics: geospatial information modeling and thematic
  mapping\/}.
\newblock Boca Raton, FL, CRC Press.

\bibitem[\protect\citename{Kang and Cressie, }2011]{kang-cressie-2011}
Kang, E.~L. and Cressie, N. (2011).
\newblock \enquote{{Bayesian inference for the Spatial Random Effects model}.}
\newblock {\em Journal of the American Statistical Association\/}, 106, 972 --
  983.

\bibitem[\protect\citename{Katzfuss, }2013]{katzfuss2013bayesian}
Katzfuss, M. (2013).
\newblock \enquote{Bayesian nonstationary spatial modeling for very large
  datasets.}
\newblock {\em Environmetrics\/}, 24, 3, 189--200.

\bibitem[\protect\citename{Katzfuss, }2017]{katzfuss2017multi}
--- (2017).
\newblock \enquote{A multi-resolution approximation for massive spatial
  datasets.}
\newblock {\em Journal of the American Statistical Association\/}, 112, 517,
  201--214.

\bibitem[\protect\citename{Katzfuss and Guinness, }2017]{katzfuss2017general}
Katzfuss, M. and Guinness, J. (2017).
\newblock \enquote{A general framework for Vecchia approximations of Gaussian
  processes.}
\newblock {\em arXiv preprint arXiv:1708.06302\/}.

\bibitem[\protect\citename{Katzfuss et~al., }2018]{katzfuss2018vecchia}
Katzfuss, M., Guinness, J., and Gong, W. (2018).
\newblock \enquote{Vecchia approximations of Gaussian-process predictions.}
\newblock {\em arXiv preprint arXiv:1805.03309\/}.

\bibitem[\protect\citename{Kleiber and Nychka, }2012]{kleiber2012nonstationary}
Kleiber, W. and Nychka, D. (2012).
\newblock \enquote{Nonstationary modeling for multivariate spatial processes.}
\newblock {\em Journal of Multivariate Analysis\/}, 112, 76--91.

\bibitem[\protect\citename{Lindgren et~al., }2011]{lindgren2011explicit}
Lindgren, F., Rue, H., and Lindstr{\"o}m, J. (2011).
\newblock \enquote{An explicit link between Gaussian fields and Gaussian Markov
  random fields: the stochastic partial differential equation approach.}
\newblock {\em Journal of the Royal Statistical Society: Series B (Statistical
  Methodology)\/}, 73, 4, 423--498.

\bibitem[\protect\citename{Liu, }1994]{liu1994collapsed}
Liu, J.~S. (1994).
\newblock \enquote{The collapsed Gibbs sampler in Bayesian computations with
  applications to a gene regulation problem.}
\newblock {\em Journal of the American Statistical Association\/}, 89, 427,
  958--966.

\bibitem[\protect\citename{Martin, }1982]{martin1982time}
Martin, W. (1982).
\newblock \enquote{Time-frequency analysis of random signals.}
\newblock In {\em Acoustics, Speech, and Signal Processing, IEEE International
  Conference on ICASSP'82.\/}, vol.~7,  1325--1328. IEEE.

\bibitem[\protect\citename{Mej{\'\i}a and Rodr{\'\i}guez-Iturbe,
  }1974]{mejia1974synthesis}
Mej{\'\i}a, J.~M. and Rodr{\'\i}guez-Iturbe, I. (1974).
\newblock \enquote{On the synthesis of random field sampling from the spectrum:
  An application to the generation of hydrologic spatial processes.}
\newblock {\em Water Resources Research\/}, 10, 4, 705--711.

\bibitem[\protect\citename{Neal, }2003]{neal2003slice}
Neal, R.~M. (2003).
\newblock \enquote{Slice sampling.}
\newblock {\em Annals of statistics\/},  705--741.

\bibitem[\protect\citename{Neto et~al., }2014]{neto2014accounting}
Neto, J. H.~V., Schmidt, A.~M., and Guttorp, P. (2014).
\newblock \enquote{Accounting for spatially varying directional effects in
  spatial covariance structures.}
\newblock {\em Journal of the Royal Statistical Society: Series C (Applied
  Statistics)\/}, 63, 1, 103--122.

\bibitem[\protect\citename{Nychka et~al., }2015]{nychka2015multiresolution}
Nychka, D., Bandyopadhyay, S., Hammerling, D., Lindgren, F., and Sain, S.
  (2015).
\newblock \enquote{A multiresolution Gaussian process model for the analysis of
  large spatial datasets.}
\newblock {\em Journal of Computational and Graphical Statistics\/}, 24, 2,
  579--599.

\bibitem[\protect\citename{Nychka, }2001]{nychka}
Nychka, D.~W. (2001).
\newblock \enquote{Spatial process estimates as smoothers.}
\newblock In {\em Smoothing and Regression: Approaches, Computation and
  Applications, rev. ed\/}, ed. M.~G. Schmiek,  393--424. New York, NY: Wiley.

\bibitem[\protect\citename{Paciorek and Schervish, }2006]{paciorek2006spatial}
Paciorek, C.~J. and Schervish, M.~J. (2006).
\newblock \enquote{Spatial modelling using a new class of nonstationary
  covariance functions.}
\newblock {\em Environmetrics\/}, 17, 5, 483--506.

\bibitem[\protect\citename{Perrin and Meiring,
  }2003]{perrin2003nonstationarity}
Perrin, O. and Meiring, W. (2003).
\newblock \enquote{Nonstationarity in ℝn is second-order stationarity in
  ℝ2n.}
\newblock {\em Journal of applied probability\/}, 40, 3, 815--820.

\bibitem[\protect\citename{Priestley, }1965]{priestley1965evolutionary}
Priestley, M.~B. (1965).
\newblock \enquote{Evolutionary spectra and non-stationary processes.}
\newblock {\em Journal of the Royal Statistical Society. Series B
  (Methodological)\/},  204--237.

\bibitem[\protect\citename{Prokhorov, }1956]{prokhorov1956convergence}
Prokhorov, Y.~V. (1956).
\newblock \enquote{Convergence of random processes and limit theorems in
  probability theory.}
\newblock {\em Theory of Probability \& Its Applications\/}, 1, 2, 157--214.

\bibitem[\protect\citename{Resnick, }2013]{resnick2013probability}
Resnick, S.~I. (2013).
\newblock {\em A probability path\/}.
\newblock Springer Science \& Business Media.

\bibitem[\protect\citename{Risser, }2016]{risser2016nonstationary}
Risser, M.~D. (2016).
\newblock \enquote{Nonstationary Spatial Modeling, with Emphasis on Process
  Convolution and Covariate-Driven Approaches.}
\newblock {\em arXiv preprint arXiv:1610.02447\/}.

\bibitem[\protect\citename{Robert, }2004]{robert2004monte}
Robert, C.~P. (2004).
\newblock {\em Monte carlo methods\/}.
\newblock Wiley Online Library.

\bibitem[\protect\citename{Sampson and Guttorp,
  }1992]{sampson1992nonparametric}
Sampson, P.~D. and Guttorp, P. (1992).
\newblock \enquote{Nonparametric estimation of nonstationary spatial covariance
  structure.}
\newblock {\em Journal of the American Statistical Association\/}, 87, 417,
  108--119.

\bibitem[\protect\citename{Sang and Huang, }2012]{sang2012full}
Sang, H. and Huang, J.~Z. (2012).
\newblock \enquote{A full scale approximation of covariance functions for large
  spatial data sets.}
\newblock {\em Journal of the Royal Statistical Society: Series B (Statistical
  Methodology)\/}, 74, 1, 111--132.

\bibitem[\protect\citename{Sayeed and Jones, }1995]{sayeed1995optimal}
Sayeed, A.~M. and Jones, D.~L. (1995).
\newblock \enquote{Optimal kernels for nonstationary spectral estimation.}
\newblock {\em IEEE Transactions on Signal Processing\/}, 43, 2, 478--491.

\bibitem[\protect\citename{Schmidt and O'Hagan, }2003]{schmidt2003bayesian}
Schmidt, A.~M. and O'Hagan, A. (2003).
\newblock \enquote{Bayesian inference for non-stationary spatial covariance
  structure via spatial deformations.}
\newblock {\em Journal of the Royal Statistical Society: Series B (Statistical
  Methodology)\/}, 65, 3, 743--758.

\bibitem[\protect\citename{Shand and Li, }2017]{shand2017modeling}
Shand, L. and Li, B. (2017).
\newblock \enquote{Modeling nonstationarity in space and time.}
\newblock {\em Biometrics\/}.

\bibitem[\protect\citename{Simpson et~al., }2017]{simpson2017penalising}
Simpson, D., Rue, H., Riebler, A., Martins, T.~G., S{\o}rbye, S.~H., et~al.
  (2017).
\newblock \enquote{Penalising model component complexity: A principled,
  practical approach to constructing priors.}
\newblock {\em Statistical Science\/}, 32, 1, 1--28.

\bibitem[\protect\citename{Spiegelhalter et~al.,
  }2002]{spiegelhalter2002bayesian}
Spiegelhalter, D.~J., Best, N.~G., Carlin, B.~P., and Van Der~Linde, A. (2002).
\newblock \enquote{Bayesian measures of model complexity and fit.}
\newblock {\em Journal of the Royal Statistical Society: Series B (Statistical
  Methodology)\/}, 64, 4, 583--639.

\bibitem[\protect\citename{Stein et~al., }2006]{stein2006spatial}
Stein, A., van~der Meer, F.~D., and Gorte, B. (2006).
\newblock {\em Spatial statistics for remote sensing\/}, vol.~1.
\newblock Springer Science \& Business Media.

\bibitem[\protect\citename{Stein, }1999]{steinML}
Stein, M.~L. (1999).
\newblock {\em Statistical Interpolation of spatial Data: Some Theory for
  Kriging\/}.
\newblock Place Springer.

\bibitem[\protect\citename{Stein, }2012]{stein2012interpolation}
--- (2012).
\newblock {\em Interpolation of spatial data: some theory for kriging\/}.
\newblock Springer Science \& Business Media.

\bibitem[\protect\citename{Stein, }2014]{stein2014limitations}
--- (2014).
\newblock \enquote{Limitations on low rank approximations for covariance
  matrices of spatial data.}
\newblock {\em Spatial Statistics\/}, 8, 1--19.

\bibitem[\protect\citename{Stein, }2015]{stein2015does}
--- (2015).
\newblock \enquote{When does the screening effect not hold?}
\newblock {\em Spatial Statistics\/}, 11, 65--80.

\bibitem[\protect\citename{Stein et~al., }2004]{stein2004approximating}
Stein, M.~L., Chi, Z., and Welty, L.~J. (2004).
\newblock \enquote{Approximating likelihoods for large spatial data sets.}
\newblock {\em Journal of the Royal Statistical Society: Series B (Statistical
  Methodology)\/}, 66, 2, 275--296.

\bibitem[\protect\citename{Tierney, }1994]{tierney1994markov}
Tierney, L. (1994).
\newblock \enquote{Markov chains for exploring posterior distributions.}
\newblock {\em the Annals of Statistics\/},  1701--1728.

\bibitem[\protect\citename{Tzeng and Huang, }2017]{TzengHuang}
Tzeng, S. and Huang, H.-C. (2017).
\newblock \enquote{Resolution adaptive fixed rank kriging.}
\newblock {\em Technometrics\/},  DOI: 10.1080/00401706.2017.1345701.

\bibitem[\protect\citename{Zammit-Mangion and Cressie, }2017]{zammit2017frk}
Zammit-Mangion, A. and Cressie, N. (2017).
\newblock \enquote{FRK: An R Package for Spatial and Spatio-Temporal Prediction
  with Large Datasets.}
\newblock {\em arXiv preprint arXiv:1705.08105\/}.

\bibitem[\protect\citename{Zhang et~al., }2018]{zhang2018smoothed}
Zhang, B., Sang, H., and Huang, J.~Z. (2018).
\newblock \enquote{Smoothed full-scale approximation of Gaussian process models
  for computation of large spatial datasets.}
\newblock {\em Statistica Sinica, accepted\/}.

\bibitem[\protect\citename{Ziemke et~al., }2005]{ziemke200525}
Ziemke, J., Chandra, S., and Bhartia, P. (2005).
\newblock \enquote{A 25-year data record of atmospheric ozone in the Pacific
  from Total Ozone Mapping Spectrometer (TOMS) cloud slicing: Implications for
  ozone trends in the stratosphere and troposphere.}
\newblock {\em Journal of Geophysical Research: Atmospheres\/}, 110, D15.

\end{thebibliography}
\nocite{*}
\end{document}